 \documentclass{emulateapj}

\usepackage{color}

\newcommand{\bb}{{\bf b}}

\newcommand{\bB}{{\bf B}}

\newcommand{\lb}{\label}
\newcommand{\be}{\begin{equation}}
\newcommand{\ee}{\end{equation}}
\newcommand{\ber}{\begin{eqnarray}}
\newcommand{\eer}{\end{eqnarray}}
\newcommand{\bers}{\begin{eqnarray*}}
\newcommand{\eers}{\end{eqnarray*}}

\newcommand{\boxi}{\hbox{\boldmath $\xi$}}

\newcommand{\Bell}{\hbox{\boldmath $\ell$}}

\shorttitle{Superdiffusion and acceleration}
\shortauthors{Lazarian \& Yan}

\begin{document}

\title{Superdiffusion of Cosmic Rays: Implications for Cosmic Ray Acceleration}

\author{A. Lazarian}
\affil{Department of Astronomy, University of Wisconsin, 475 North Charter Street, Madison, WI 53706, USA}

\author{Huirong Yan}
\affil{KIAA, Peking University, 5 Yi He Yuan Rd, Beijing, 100871, China}

\begin{abstract}

Diffusion of cosmic rays (CRs) is the key process of understanding their propagation and acceleration. We employ the description of spatial separation of magnetic field lines in MHD turbulence in Lazarian \& Vishniac (1999) to quantify the divergence of magnetic field on scales less than the injection scale of turbulence and show this divergence induces superdiffusion of CR in the direction perpendicular to the mean magnetic field. The perpendicular displacement squared increases, not as distance $x$ along magnetic field, which is the case for a regular diffusion, but as the $x^{3}$ for freely streaming CRs. The dependence changes to $x^{3/2}$ for the CRs propagating diffusively along magnetic field. In the latter case we show that it is important to distinguish the perpendicular displacement in respect to the mean field and to the local magnetic field. We consider how superdiffusion changes the acceleration of CRs  in shocks and show how it decreases efficiency of the CRs acceleration in perpendicular shocks. We also demonstrate that in the case when small-scale magnetic field is being generated in the pre-shock region, an efficient acceleration can take place for the CRs streaming without collisions along magnetic loops. 
\end{abstract}

\keywords{ galaxies: magnetic fields -- methods: theoretical -- MHD -- turbulence}

\section{Introduction}

Cosmic ray (CR) acceleration and diffusion are the long standing problems with a lot of literature describing the processes \citep[see monographs by][ and references therein]{Ginzburg74_book3, Schlickeiser02}. Turbulence plays the key role for both processes and as a result, advances in understanding magnetohydrodynamic (MHD) turbulence advance both fields.

Recent decades have been marked by substantial progress in understanding MHD turbulence and this has has already shifted some of the cosmic ray paradigms. For instance, the discovery of the scale-dependent anisotropy of Alfvenic turbulence \citep[][henceforth GS95]{GS95}, isotropy of fast modes \citep{CL02_PRL, CL03} as well as the quantitative studies of mode coupling (GS95, \citealt{CL02_PRL}) resulted in identifying of fast modes as the major scattering agent in the typical ISM conditions \citep{YL02,YL04}. 

In this paper we explore the consequences of another advance of MHD turbulence theory, namely, the understanding of the magnetic line separation in Alfvenic turbulence \citep[][henceforth LV99]{LV99}, which presents a close analog of the separation of particles in turbulent media due to the well known process of  Richardson diffusion. This type of explosive growth of the separation between the particles, i.e. as $(time)^{3/2}$, on the scales less than the turbulence injection scale $L$ was inferred from fluids experiments many decades ago \citep[see][]{Richardson}. The turbulent wandering of magnetic fields was quantified in LV99 providing the separation of magnetic field lines increasing as $(distance)^{3/2}$, where the distance is measured along the magnetic field lines. The intimate connection of this to the Richardson diffusion was revealed in \citet[][henceforth ELV11]{ELV11}. The LV99 expression for the field wandering was also numerically confirmed  \citep[see][]{LVC04, Maron04} and was employed for the description magnetized plasma thermal conduction (Narayan \& Medvedev 2001, Lazarian 2006) and cosmic ray propagation \citet[][henceforth YL08]{YL08}. A recent numerical confirmation of the Richardson diffusion for magnetized fluids can be found in Eyink et al. (2013)\footnote{We may not parenthetically that the Richardson diffusion in
magnetized fluid also demonstrates the violation of flux freezing in turbulent fluids of arbitrary conductivity. This is related to the theory of fast turbulent
reconnection presented in LV99 to which the numerical study in Eyink et al. (2013) provides an additional testing (see also Kowal et al. (2009, 2012).}. 

In what follows we will refer to the magnetic field divergence on scales less than the injection scale $L$ as the {\it spatial Richardson diffusion}
or, when it does not cause confusion, simply as the {\it Richardson diffusion}. The intrinsic relation between the magnetic field spatial and time super-diffusion is demonstrated and discussed in detail in ELV11 and this justifies the use of the same term for the two closely related processes. 

The spatial Richardson diffusion substantially changes the perpendicular diffusion for CRs both streaming and diffusing along magnetic field lines and this entails important consequences for the cosmic ray transport and acceleration. The effect of the Richardson diffusion is important as  the difference in the acceleration efficiency of parallel and perpendicular shocks has been the subject of intensive discussions in the literature, e.g. see \citep{Jokipii87}. 
In the paper we show that the superdiffusion of cosmic rays arising from the Richardson diffusion of magnetic field lines can substantially modify the arguments.

The notion of the superdiffusive behavior can be traced back to the paper by \citet{Jokipii73} as well as that by \citet{Skilling74}. There the fast deviations of the magnetic  field lines was reported in the context of the problem of cosmic ray diffusion.The quantitative results in the aforementioned
papers are, as we discuss further in the paper, inconsistent with the prediction of Richardson diffusion. More recently, superdiffusion was reported in the analysis of solar wind data \citep[see][]{Perri2009} where the phenomenon was attributed to cosmic rays ballistic behavior or Levi flights. In contrast, in our study we do not appeal to the hypothetical Levi flights, the nature of which for CRs is not clear. 

This paper is a continuation of our exploration of the perpendicular diffusion of CRs. In our earlier paper, namely Yan \& Lazarian (2008, henceforth YL08), we mostly dealt with cosmic ray diffusion on the scales larger than the injection scale $L$. There we also had a short discussion of particle transport on scales less than $L$ for which we considered scale-dependent diffusion coefficients, which, as we argue here, reflect superdiffussive behavior. The predictions in YL08 have been confirmed by numerical simulations that employed results of 3D MHD simulations \citep{Xu_Yan}. We discuss the results of the latter testing in view of our theoretical study within this work. 

The transport of CRs at the scale less than $L$ that we deal in the paper is an important regime for many astrophysical applications. For instance, for interstellar media $L\sim 100$~pc (see Elmegreen \& Scalo 2004, Chepurnov et al. 2010), it is around 50 pc for M51 \citep{Fle11}, and about 20 pc in the fan region in the outerskirts of the Galaxy \citep{Iac13}. It is generally accepted that the acceleration processes in shocks happen on the scale comparable or smaller than that. While we show that superdiffusive behavior can alter some of the popular ideas on CR acceleration in shocks, we also discuss a process that avoids the limitations of superdiffusive behavior. This process is related to the acceleration of CRs streaming along small scale magnetic generated in the pre-shock and post-shock regions. Within this process the effective mean free path of the particles is determined by
the entangled magnetic field structure (see Lazarian 2006) rather than scattering of particles at magnetic field perturbations.

In what follows, we discuss in \S 2 some basis properties of MHD turbulence which determines particle transport,  the Richardson diffusion
of magnetic fields in \S 3, briefly discuss the process of subdiffusion in \S 4. The modifications of the shock acceleration in the presence of superdiffusion are dealt with in \S 5. \S 6 provides a quantitative study of the effects of superdiffusion for a few idealized astrophysically motivated settings. In \S 7 we discuss a process of CR acceleration while they stream along the small-scale magnetic field generated in the pre-shock and post-shock regions.We show that a very fast and efficient acceleration is possible. The  discussion of our findings and summary are provided in \S 8 and \S 9 respectively.

\section{MHD turbulence as the key factor for cosmic ray propagation}
\label{MHD}

It is generally accepted that CRs follow magnetic field lines and get scattered by magnetic perturbations. The statistics of magnetic field lines and the nature of perturbations are determined by magnetic turbulence. Therefore it is essential to use the model of turbulence that has solid theoretical foundations and agrees with the results of numerical simulations for describing cosmic ray propagation and acceleration\footnote{A frequently used in magnetospheric and heliospheric research model of Alfvenic turbulence is so-called "slab + 2D" model \cite{Bieber88} was introduce to empirically represent scattering of energetic particles in Solar wind. This model is not confirmed by numerical simulations and presents an approximate empirical treatment of a particular turbulent system. As we discuss in this section, the existing numerical simulations instead support GS95 theory and its compressible MHD extensions.} 

 Alfvenic perturbations for decades constituted the textbook default for describing propagation of cosmic rays. While this changed as fast modes were
 identified as the major agent for resonance scattering \citep{YL02, YL04} in this work we confirm their major role for
 diffusion perpendicular to the mean field\footnote{The inefficiency of Alfven modes stems from two factors. One of them is a scale-dependent anisotropy that makes perturbations extremely elongated at small scales if the energy is injected, as it usually happens in astrophysical systems, e.g. the ISM or intracluster gas, at large scales. The other is the rapid decrease of the energy in terms of $k_{\|}$, where $\|$ is measured in the direction of local magnetic field at the scale of the Larmor radius of the resonance particle. Those factors are discussed in \citet{Chandran00} and \citet{YL02}.}.
 
The possibility to discuss Alfven modes separately is based on both numerical and theory arguments. The numerical study in \citet{CL02_PRL} demonstrated that in compressible MHD turbulence the Alfvenic modes develops an independent cascade which is marginally affected by the fluid compressibility. This observation corresponds to theoretical expectations of the GS95 theory that we briefly describe below \citep[see also][]{LG01,CL02_PRL}. The corresponding numerical studies of the decomposition of MHD turbulence
into Alfven, slow and fast modes is performed in \citet{CL03} and \cite{KL10} with the Fourier technique and wavelets, respectively.

The theory of MHD turbulence has become testable due to the advent of numerical simulations \citep[see][]{Biskampbook}
that confirmed (see \citealt{LBYO, Lazarian_rev12}, Brandenburg \& Lazarian 2013 and references therein) the general expectation of magnetized Alfvenic eddies being elongated in the direction of magnetic field \citep[see][]{Shebalin83, Higdon84} and provided results in agreement with the 
quantitative predictions for the variations of eddy elongation obtained  in GS95.

\begin{table*}
\caption{Regimes of MHD turbulence and magnetic diffusion}
%\vskip4mm
\centering
%\begin{tabular}{column = lcr}
\begin{tabular}{cccccccc} 
\multicolumn{8}{}{{\bf Table 1}} \\
\multicolumn{8}{}{Regimes and types of magnetic diffusion for MHD turbulence} \\
\hline
\hline
Type                       & Injection  &  Range                & Spectrum   &Motion & Ways & Magnetic & Squared separation\\
of MHD turbulence & velocity  & of scales              & E(k)           &   type & of study& diffusion  &  of lines \\
\hline
Weak                      & $V_L<V_A$ & $[l_{trans}, L]$   & $k_{\bot}^{-2}$ & wave-like & analytical & diffusion & $\sim sLM_A^2$ \\
\hline
Strong         &                   &                                  &                            &   anisotropic &               &                     &\\
subAlfvenic& $V_L<V_A$ & $[l_{min}, l_{trans}]$ &  $k_{\bot}^{-5/3}$ & eddy-like & numerical & Richardson &  $\sim \frac{s^3}{L} M_A^4$\\
\hline
Strong             &                    &                   &                                      &     isotropic                           & &                 &  \\
superAlfvenic & $V_L> V_A$ & $[l_A, L]$ &        $k_{\bot}^{-5/3}$        &  eddy-like & numerical & diffusion &     $\sim s l_A$       \\
\hline
Strong           &                      &                   &                                          &  anisotropic    &             &           &     \\
superAlfvenic & $V_L> V_A$ & $[l_{min}], l_A$ &   $k_{\bot}^{-5/3}$   & eddy-like &     numerical & Richardson &  $\sim \frac{s^3}{L}M_A^3$\\
\hline
& & & &&&&\\
\multicolumn{8}{l}{\footnotesize{$L$ and $l_{min}$ are the injection and perpendicular dissipation scales, respectively. $M_A\equiv \delta B/B$, $l_{trans}=LM_A^2$ for $M_A<1$  and $l_{A}=LM_A^{-3}$.}}\\
\multicolumn{8}{l}{\footnotesize{for $M_A<1$. For weak Alfvenic turbulence $\ell_{\|}$ does not change. $s$ is measured along magnetic field lines.}}\\
\end{tabular}
\end{table*}

The hydrodynamic counterpart of the MHD turbulence theory is the famous \citet{Kolmogorov} theory of turbulence. In the latter theory energy is injected at large scales, creating large eddies which correspond to large $Re$ numbers and therefore do not dissipate energy through viscosity\footnote{Reynolds number $Re\equiv L_fV/\nu=(V/L_f)/(\nu/L^2_f)$ which is the ratio of an eddy turnover rate $\tau^{-1}_{eddy}=V/L_f$ and the viscous dissipation rate $\tau_{dis}^{-1}=\eta/L^2_f$. Therefore large $Re$ correspond to negligible viscous dissipation of large eddies over the cascading time $\tau_{casc}$ which is equal to $\tau_{eddy}$ in Kolmogorov turbulence.} but transfer energy to smaller eddies. The process continues until the cascade reaches the eddies that are small enough to dissipate energy over eddy turnover time. In the absence of compressibility the hydrodynamic cascade of energy is $\sim v^2_\ell/\tau_{cas, \ell} =const$, where $v_\ell$ is the velocity at the scale $\ell$ and the cascading time for the eddies of size $\ell$ is $\tau_{cas, \ell}\approx \ell/v_\ell$. From this the well known relation $v_\ell\sim \ell^{1/3}$ follows.

 It is easy to see why magnetic turbulence is anisotropic. One can imagine eddies mixing magnetic field lines perpendicular to the direction of local magnetic field. For such eddies the original Kolmogorov treatment is applicable resulting in perpendicular motions scaling as $v_\ell\sim \ell_{\bot}^{1/3}$, where $\ell_{\bot}$ denotes eddy scales measured perpendicular to magnetic field. These mixing motions induce Alfvenic perturbations that determine the parallel size of the magnetized eddy.  The key stone of the GS95 theory is {\it critical balance}, i.e. the equality of the eddy turnover time $\ell_{\bot}/v_\ell$ and the period of the corresponding Alfven wave $\sim \ell_{\|}/V_A$, where $\ell_{\|}$ is the parallel eddy scale and $V_A$ is the Alfven velocity. Making use of the earlier expression for $v_\ell$ one can easily obtain $\ell_{\|}\sim \ell_{\bot}^{2/3}$, which reflects the tendency of eddies to become more and more elongated as the energy cascades to smaller scales.

It is important to stress that the scales $\ell_{\bot}$ and $\ell_{\|}$ are measured in respect to the system of reference related to the direction of the local magnetic field ``seen'' by the eddy. This notion was not present in the original formulation of the GS95 theory and was added to it later \citep{LV99, CV00, MG01,CLV_incomp}. In terms of mixing motions, it is rather obvious that the free Kolmogorov-type mixing is possible only in respect to the local magnetic field of the eddy rather than the mean magnetic field of the flow\footnote{Fast reconnection is also required for the mixing motions. The 
reconnection theory in LV99 provides the necessary rates to make the GS95 picture of turbulence self-consistent.}.

The quantitative properties of Alfvenic turbulence that we mostly deal with can be expressed for subAlfvenic turbulence using LV99 expressions:
\begin{equation}
\ell_{\|}\approx L_i \left(\frac{\ell_{\bot}}{L_i}\right)^{2/3} M_A^{-4/3},
\label{Lambda}
\end{equation}
\begin{equation}
\delta u_{\ell}\approx u_{L} \left(\frac{\ell_{\bot}}{L_i}\right)^{1/3} M_A^{1/3},
\label{vl}
\end{equation} 
which, unlike those in GS95, are valid for the arbitrary injection velocity $<V_A$. For subAlvenic turbulence the ratio $M_A$ 
is
\be
M_A\equiv \frac{V_L}{V_A}=\frac{\delta B}{B}<1,
\ee
while when the injection is superAlfvenic $M_A=V_L/V_A>1$. 
 
We feel that the claims that the $-5/3$ slope predicted by the GS95 is incorrect are not substantiated. For instance, recent work by \citep{BL10} shows that present day numerical simulations are unable to reveal the actual inertial range of MHD turbulence making the discussions of the discrepancies of the numerically measured spectrum and the GS95 predictions rather premature\footnote{More recent higher resolution simulations by \cite{Bere11} reveal the predicted $-5/3$ spectral slope. In any case, the proposed additions to the GS95 model do not change the nature of the physical processes that we discuss below.}.
 
GS95 theory assumes the isotropic injection of energy at scale $L$ corresponding to the Alfven Mach number
\begin{equation} 
M_A\equiv \delta B/ B =1, 
\end{equation}
where $\delta B$ is the field fluctuation and $B$ is the mean field. For the incompressible MHD turbulence $\delta B/B$ is equal to $V_L/V_A$, where
$V_L$ is the injection velocity and the Alfv\'en velocity in
the fluid is $v_A$. Thus it provides the description
of transAlfvenic turbulence with $V_L=v_A$. This model was later generalized
for both subAlfvenic, i.e. $M_A<1$, and superAlfvenic, i.e. $M_A>1$, cases (see \citealt[][henceforth LV99]{LV99}, \citealt[][respectively]{Lazarian06}; see also
Table~1).  Indeed, if $M_A>1$, instead of the driving scale $L$ for  one can use another scale,
namely $l_A=LM_A^{-3}$, which is the scale at
which the turbulent velocity equals to $V_A$ (see Lazarian 2006). The scale $l_A$ is termed Alfvenic
scale (Lazarian 2006) and is the scale
below which magnetic field lines become stiff and their back-reaction becomes essential. For $M_A\gg 1$
magnetic fields are not dynamically important at the scales larger than $l_A$ and the
turbulence at those scales follows the isotropic
 Kolmogorov cascade $\delta u_\ell\sim \ell^{1/3}$ over the
range of scales $[L, l_A]$. 
At the same time, if $M_A<1$, the turbulence obeys GS95 scaling (also called ``strong''
MHD turbulence) not from the scale $L$, but from a smaller scale $l_{trans}=LM_A^2$, while in the range $[L, l_{trans}]$ the turbulence is ``weak''. 

The properties of weak and strong turbulence are rather different. Weak turbulence is wave-like turbulence with wave packets undergoing many collisions before transferring energy to small scales.  Weak turbulence, unlike the strong one, allows an exact analytical treatment \citep{Galtier2001}. On the contrary, the strong turbulence is eddy-like with cascading happening similar to Kolmogorov turbulence within roughly an eddy turnover time. The strong interactions between wave packets prevent the use of perturbative approach and do not allow exact derivations. It were the numerical experiments that proved the predicted scalings for incompressible MHD turbulence \citep[see][]{CV00, MG01,CLV_incomp, BL10,Bere11} and for the Alfvenic component of the compressible MHD turbulence \citep{CL02_PRL, CL03,KL10}. 

One also should keep in mind that the notion ``strong'' should not be associated with the amplitude of turbulent motions but only with the strength of the non-linear interaction. As the weak turbulence evolves, the interactions of wave packets {\it get stronger} transferring the turbulence into the strong regime. In this case, the amplitude of the perturbations can be very small.

The theories above assume an isotropic injection of turbulence at large scales. This simplest type of energy injection
can happen in both media with and without a mean magnetic field (see simulations of Cho \& Lazarian (2003) with a mean magnetic
field). Isotropic injection of turbulent energy is also assumed in our paper. An anisotropic injection of energy can happen e.g. due to cosmic ray interaction 
with compressible turbulence \citep[e.g.][]{LB06, YL11}. In this situation, the energy is injected at the cosmic rays gyroradii and in the form of plane waves. These waves will interact and get cascaded by the external Alfvenic turbulence \citep{YL02, FG04, BL_wave}. Magnetic reconnection provides another example of anisotropic turbulence driving (see Lazarian et al. 2013), but this is not expected to be the dominant source of turbulent
energy in the galaxy.  

Astrophysical turbulence presents a wide variety of conditions. MHD turbulence in solar 
wind (Leamon et al. 1998) is different from the interstellar turbulence (see Big Power Law in Armstrong et al 1994 and Chepurnov \& Lazarian 2010, the measured spectra of velocity turbulence 
in HI and CO are reviewed e.g. in Lazarian 2009) both in terms of the injection scales and the injection velocities. The media are also different in terms of its magnetization (see reviews by Elmegreen \& Scalo 2004, McKee \& Ostriker 2007, \citealt{BrL13} and references therein). This induces substantial variations of the properties of turbulence including the variations of its Alfvenic Mach number $M_A$. Determining this parameter from observations is not trivial, as this number depends on the ratio of magnetic fluctuation to the mean magnetic field at the injection scale. The value of this scale is still the subject of controversy, but we believe
that for interstellar turbulence it should be determined by the supernovae injection and be of the order of 100 pc. 

Radio continuum polarization observations find that in external galaxies on the scale of several hundred parsec the ordered magnetic field is 1-5 $\mu$G, while the total magnetic field is of
the order of 9 $\mu$G (Beck \& Wielebinski 2013). The values of the mean and chaotic field that we need to calculate $M_A$ are difficult to obtain from these input data, as the measurements are not done at the turbulence injection scale $L$, which is smaller. The ratio of the chaotic to the total magnetic field on scales larger than $L$ depends on the action of magnetic dynamo, which remains a theory facing many challenges, e.g. related to the helicity conservation (see Vishniac \& Cho 2001). We feel that the existing data is compatible with $M_A$ of the order of unity, which corresponds to the energy equipartition of between turbulence and magnetic field. Naturally, better determination of $L$ and the magnetic field fluctuations on the scale $L$ are required.

\section{Richardson diffusion of magnetic field and cosmic rays}

Above the turbulent injection scale $L$, naturally, magnetic field lines undergo random walk. In what follows 
we focus on the field line divergence over scales less than $L$ where the phenomenon of Richardson diffusion takes place.
 These are important scales for cosmic ray propagation and
acceleration, as, for instance, the energy injection scale in the interstellar medium is around 100 pc \citep[see][]{Chep2010}. Nevertheless our treatment below is quite general and it covers various astrophysical
 environements. Therefore we consider a variety of possible situations for subAlfvenic, superAlfvenic turbulence and for the situations when the mean free path of CR is larger or smaller than the injection scale. 

\subsection{Magnetic field wandering}
\lb{B_wander}
Turbulence is frequently visualized as a hierarchy of interacting eddies. In the well-known example of Kolmogorov turbulence with the cascading
rate $\epsilon\approx v_\ell^3/\ell=const$ the relative velocities of particles increase as the scale of motions increases according to the famous $1/3$ law, i.e.
$v(\ell)\sim v_0(\ell/L)^{1/3}$, where $v_\ell$ is the difference of velocities of particles separated by distance $\ell$, $v_0$ is the injection velocity at the scale $L$. As $v_\ell\sim \frac{d}{dt} \ell(t)$ one can write 
\begin{equation}
\frac{d}{dt} \ell(t)\sim(\epsilon \ell)^{1/3}
\end{equation}
which gives solution $\ell^2\sim \epsilon t^{3}$, which corresponds to the famous Richarson law \citep{Richardson}  of particle diffusion in hydrodynamic turbulence\footnote{Note, that the Richarson law was discovered empirically prior to the formulation of the Kolmogorov theory.}. An interesting feature of this law is the accelerated separation of fluid particles, which arises from the fact that as time goes on the particles become enter eddies of larger size which, according to the Komogorov theory, have larger velocity dispersions.

In the case of magnetized flows the Richardson diffusion represents itself in terms of magnetic field wandering\footnote{This field wandering is closely related to the fast magnetic reconnection in a turbulent flow (LV99), which in its turn is related to the violation of magnetic flux freezing in turbulent fluids (ELV11). Indeed, it is easy to see that the wandering of magnetic field lines related to turbulent motions is impossible without fast reconnection and the perpetual changing of magnetic field topology. Without such a reconnection magnetic field lines in turbulent fluid would produce a felt-like structure arresting all hydrodynamic type mixing motions. Although a picture of magnetic field wave-like turbulence with waves propagating in the elastic structure of magnetic interlocked field lines is possible to imagine, it corresponds neither to numerical simulations nor to observations of turbulent diffuse interstellar media, molecular clouds or solar wind \citep[see][]{Armstrong95, Leamon98, ElmegreenScalo, McKee_Ostriker2007, Laz09rev}. As we 
mentioned earlier, fast LV99 reconnection makes GS95 picture self-consistent enabling magnetic field lines to change their topology over one eddy turnover time. } (LV99). LV99 quantified the process and its relation to the Richardson diffusion was established in ELV11. 
%\bibitem[Leamon et al.(1998)]{1998JGR...103.4775L} Leamon, R.~J., Smith, 
%C.~W., Ness, N.~F., Matthaeus, W.~H., \& Wong, H.~K.\ 1998, \jgr, 103, 4775

%The theory of magnetic field wandering within GS95 turbulence was presented in LV99 where it was established that the line separation grows with the distance along magnetic field $s^{3/2}$ presents an analog of the Richardson law. This behavior is closely related to the time-dependent Richardson
%separation of particles of magnetized fluid as was shown in ELV11. Thus we use the term Richardson diffusion for both time dependent behavior of particles and magnetic field line separation. The Richardson diffusion in terms of magnetic field divergence was confirmed in numerical simulations \citep[see][]{LVC04}, while the Richardson-type time separation of the magnetic field lines was demonstrated in \cite{EV13_nature}. This law was shown to be important both for magnetic reconnection (LV99, ELV11), for heat transfer \citep{NM01, Lazarian06} and cosmic ray diffusion (YL08). In this paper we also show how it modifies the acceleration of cosmic rays.

\subsection{CR streaming along wandering magnetic field lines}

Magnetic field scattering by Alfvenic turbulence may be very inefficient if the energy is injected at scales much larger
than the gyroradius of the CR (Chandran 2000, Yan \& Lazarian 2002). This is due to the scale dependent anisotropy of the Alfvenic waves. Thus in the absence of the fast modes that were identified as the main scattering agent in most astrophysical environments, e.g. the interstellar medium (Yan \& Lazarian 2002), the scattering may be very inefficient and CRs may freely stream along magnetic field lines. Fast modes can be depleted at small scales due to dampling, e.g. the collisionless damping (see quantitative study in Yan \& Lazarian 2004).

If the mean free path of a CRs $\lambda_{CR}$ is much larger than the scale over which CR propagation is considered, the CRs stream along magnetic field and their diffusion perpendicular mean magnetic field is determined by the divergence of magnetic field lines. In this section we quantify this type of separation.  

 First we consider the  regime of Alfven Mach numbers $M_A<1$, which is important for many astrophysical implications. We describe the divergence of field lines that start at points displaced by vector $\Bell$. The displacement increases  is considered as one moves a distance $s$ in arc-length along the field lines passing through the two points. Below we use a more formal approach that was suggested in ELV11 and which is complementary to an
 intuitive approach adopted in LV99. Within the former approach, the corresponding equation describing the change in separation is
\be 
{{d}\over{ds}} \Bell(s) =\hat{\bb}(\boxi'(s))-\hat{\bb}(\boxi(s)), 
\ee
where $\Bell(s) =\boxi'(s)-\boxi(s),$ and $\hat{\bb}=\bB/|\bB|$ is the unit tangent vector along the magnetic field-line.  

We leave the details of the derivation to the Appendix A, while here we provide the results. For strong Alfvenic turbulence
we get
\be 
{{d}\over{ds}}\ell_\perp^2\sim D_\perp^B(\ell)\sim (\delta u_\ell/v_A)^2 \ell_\| \sim L\left({{\ell_\perp}\over{L}}\right)^{4/3}
M_A^{4/3}, 
\label{diff_strong}
\ee
while for the weak turbulence regime one should use 
$\ell_\|=L$ (see \S 2 and Table~1) being a constant, which  gives 
\be  
{{d}\over{ds}}\ell_\perp^2\sim D_\perp^B(\ell) \sim L M_A^4,
\label{diff_weak}
\ee
or
\be
\ell_\perp^2\sim sLM_A^4.
\label{diff_weak2}
\ee

The predicting behavior was tested numerically. Figure \ref{lazarianetal} illustrates both the Richardson diffusion of magnetic field lines and its transition to the ordinary diffusion at the scale larger than
  the energy injection one. The results confirm the Richardson scaling or magnetic line separation that follows from Eqs. (\ref{diff_strong}) and (\ref{bot_strong}). Similar numerical results were also obtained in Maron et al. (2004). We note that for the case of $\lambda_{CR}\gg L$ cosmic ray trajectories trace magnetic field lines and therefore the divergence of
CR trajectories is identical to that of magnetic field.
  
  \begin{figure}
  \includegraphics[width=8cm]{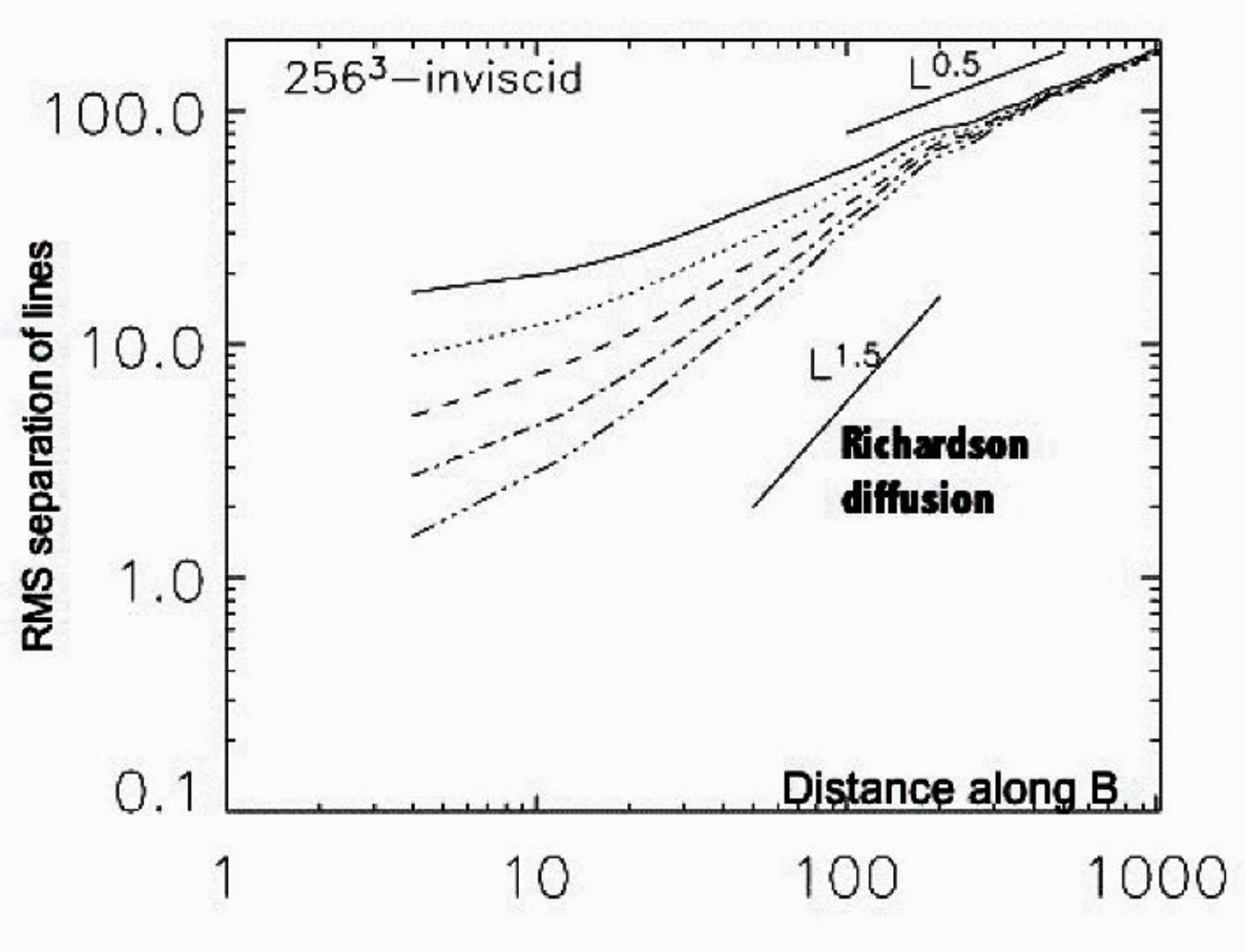}
    \caption{Magnetic field separations in MHD driven simulations. Cosmic rays
that stream along magnetic fields without scattering experience the same
saparation. Two regimes, the Richardson one at small scales (see Eq. (\ref{bot_strong})) and the
    ordinary random walk diffusion of magnetic field lines at large scales are seen.  Modified from Lazarian et al. (2004).}
  \lb{lazarianetal}
\end{figure}

Thus we clearly observe two different regimes. For the weak turbulence the solution of Eq. (\ref{diff_weak}) induces a usual random walk diffusion
law, while solving Eq. (\ref{diff_strong}) provides the Richardson-type scaling of magnetic field line separation with
$\ell_\perp^2\sim s^3$.  The Richardson law in terms of magnetic field separation in time was
confirmed with numerical simulations involving many time frames of the driven MHD turbulence  \citep{EV13_nature}. The Richardson diffusion of magnetic field lines was also observed through tracing of CR that ballistically follow turbulent magnetic field (Xu \& Yan 2013). 

The solution of Eq. (\ref{diff_strong}) in terms of $\ell_{\bot}$ dependence on $s$ is
\be
\ell_{\bot}^2 \sim \frac{s^3}{27L} M_A^4,
 \label{bot_strong}
 \ee
 where we stress the importance of the $M_A^4$ dependence, which contrasts with the $M_A^2$ dependence in the classical studies \citep[see][] {Jokipii_Parker1969}\footnote{The scaling given by Eq. (\ref{bot_strong}) can also be obtained by observing that
$$
\frac{d}{ds}\ell_\bot \simeq \frac{\delta b_\ell}{B_0}\sim \frac{\delta u_\ell}{v_A}.
$$
Inserting the turbulent velocity given by Eq.~(\ref{vl}),  one gets
$$ \ell_\bot^2 \sim (s^3/L_i) M_A^4  $$.}.
This dependence translates in the corresponding $M_A^4$ dependences for the perpendicular diffusion of CRs, which means a much stronger suppression of perpendicular diffusion by magnetic field. The vivid feature of Eq. (\ref{bot_strong}) is that $\ell_{\bot}<s$, as for the Richardson diffusion it is required that scales $s$ are less than the injection scale $L$. 
 
  \begin{figure}
        \includegraphics[width=0.45\textwidth]{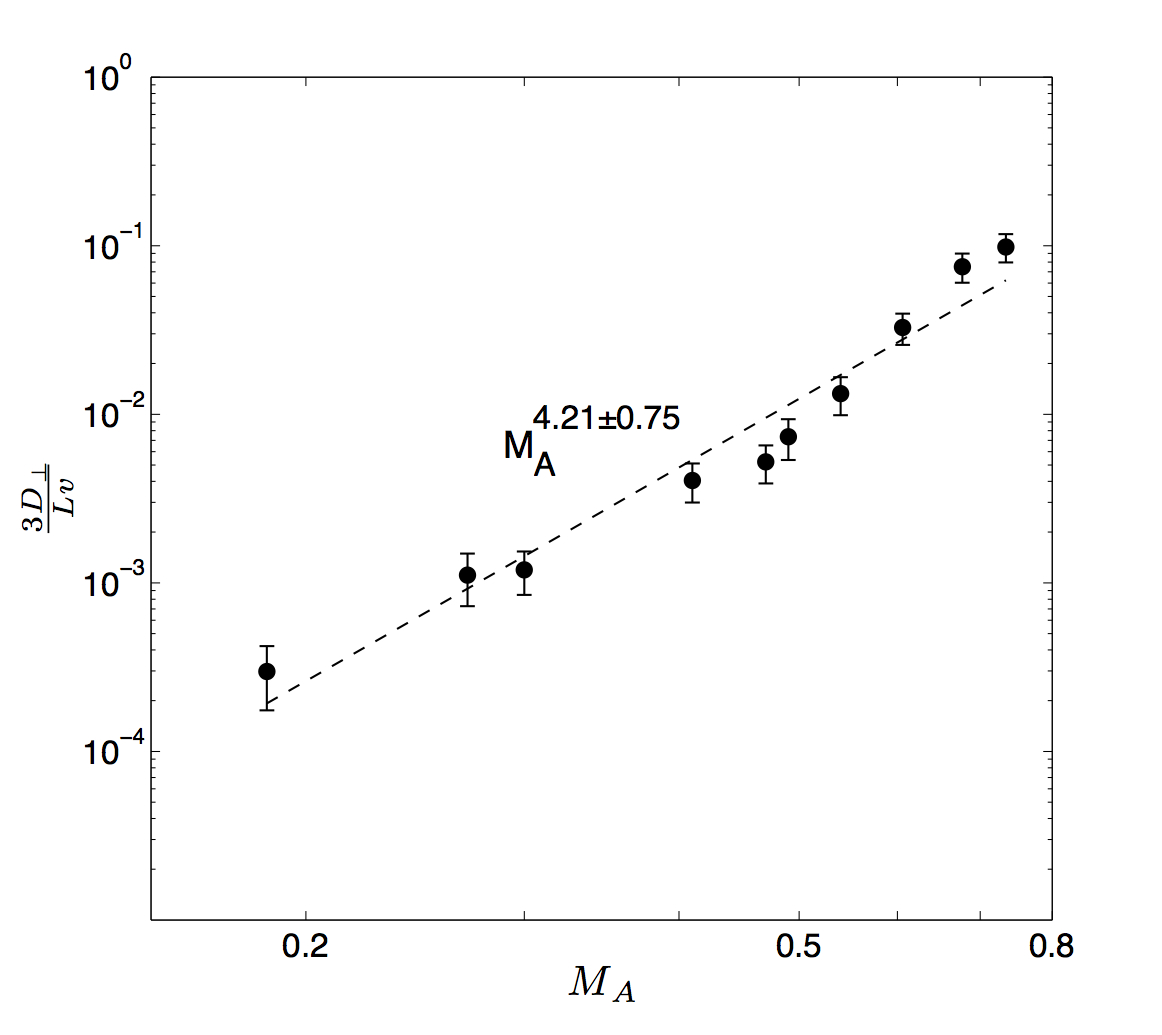}
     \caption{Numerical simulations of CR propagation along wandering magnetic field in case of free streaming. The perpendicular diffusion coefficient normalized by particle mean parallel speed $v/3$ multiplied by turbulence injection scale $L$ vs. $M_A$. From \citet[][]{Xu_Yan}.}
     \label{MA4th}
 \end{figure}
 
 Figure \ref{MA4th} illustrates the results of numerical simulations of the propagation of CRs in compressible MHD turbulence in Xu \& Yan (2013). Unlike many other CR studies, these results are obtained with turbulence produced by direct numerical simulations with the MHD compressible code (see more details on the code in Cho \& Lazarian 2003). Therefore these results correspond to the propagation of CRs in GS95-like turbulence\footnote{It is rather difficult to generate turbulence with the scale-dependent anisotropy. Thus the simulations that use synthetic turbulence cubes without this feature are missing essential physics.} and they agree well with the theoretical expectations given by Eq. (\ref{bot_strong}). We can clearly see that the simulations support the $M_A^4$ dependence of the perpendicular coefficient instead of the generally accepted $M_A^2$ one. 

 \begin{figure}
        \includegraphics[width=0.48\textwidth]{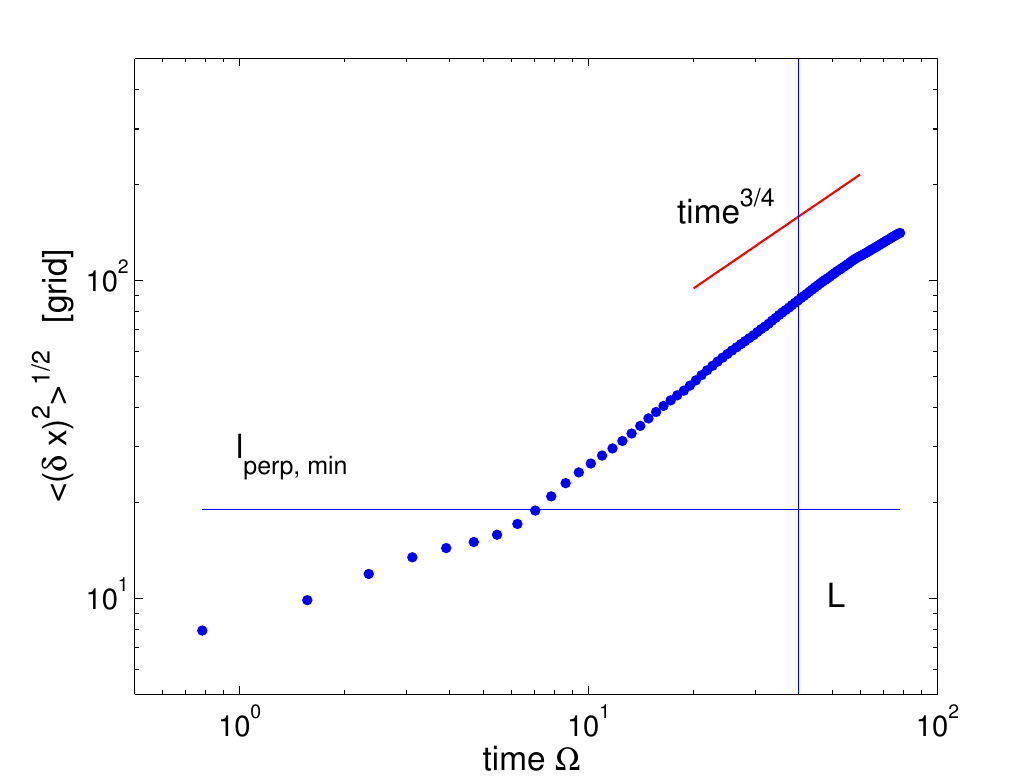}
     \caption{Numerical simulations of CR propagation. Superdiffusion of CRs in the regime of high rate scattering
as confirmed in numerical simulations (see also eq.\ref{bot_diff1}). From \citet[][]{Xu_Yan}. }
     \label{MA4th_b}
 \end{figure}
   
  The perpendicular displacement given by Eq. (\ref{diff_weak}) is an obvious random walk in terms of the path along the magnetic field lines
  \be
  \ell_{\bot}^2\sim s L M_A^4,~~~~M_A<1,
  \label{bot_weak}
  \ee
  where the dependence on $M_A^4$ is prominently present. In weak turbulence the parallel scale does not change (see LV99, Galtier 2001) and therefore it is always equal to the injection scale. The change happens when the turbulence reaches the scale $l_{trans}$ given in Table~1 where  at which it transfers to the strong turbulence. In astrophysical situations the there can be more than one sources of weak turbulence injection\footnote{Cosmic ray instabilities, e.g. streaming instability produces
Alfvenic waves with wave vector parallel to the local magnetic field direction. As these waves are being reflected of density inhomogeneities this creates oppositely moving Alfven waves which entails an imbalanced weak cascade \citep[see][]{LG01}. These inhomogeneities may dominate
at sufficiently small scales, which makes the discussed regime of CR diffusion astrophysically relevant.}.  Therefore in Eq. (\ref{bot_weak}) $s\gg L$ and this regime corresponds to the usual random walk.
  
For Alfv\'en Mach numbers larger than unity, i.e. $M_A>1$, at  scales larger than the Alfvenic scale $l_A$ given in Table 1, the dynamics of magnetic field is dominated by hydrodynamic turbulence. However, the magnetic field is entangled at the scale $l_A$ and thus the separation of magnetic field lines is a random walk process with the step $l_A$. Therefore the mean squared separation between the magnetic field lines $l_{\bot}^2$ is increasing with the distance tracked along the magnetic field line $s$ as 
\begin{equation}
l_{\bot}^2\sim s l_A~~~.
\end{equation}
As a result, for the scales $[ l_A, L]$ magnetic field lines undergo diffusion and the transport of cosmic rays that stream along magnetic field is diffusive on the scales larger than $l_A$.

For scales smaller than $l_A$ the turbulence is of GS95-type with the injection scale of $L=l_A$ and the injection velocity $V_L=V_A$. The
 separation of magnetic field lines can be obtained easily from Eq. (\ref{bot_strong}):
 \be
\ell_{\bot}^2 \sim \frac{s^3}{27l_A}\sim \frac{s^3 }{27 L} M_A^3,~~~M_A>1~~[l_{min}, l_A],
 \label{bot_super}
 \ee
which the $M_A^3$ dependence reflects the fact that the GS95 cascade starts at $l_A$ rather that
at the injection scale of turbulence $L$. Otherwise, all our arguments about magnetic field line divergence for the case of $M_A<1$ are applicable for the scales $l<l_A$ provided that the scale $l_A$ is used instead of $L$. The physical basis for this is that the magnetic fields at scale less that $l_A$ get stiff and are not pliable to being bend by hydrodynamic turbulence. 

Streaming CR\footnote{Here streaming is understood in terms
of motion without scattering. Note that when cosmic rays stream in
one direction they may produce ``streaming instability''. But we do 
not consider such a case.} move along magnetic field lines with the velocity $c\mu$, where $\mu$ is the cosine of pitch angle.
If we do not consider cosmic rays with the pitch angle close to 90 degrees, the velocity of streaming is of the order
of the velocity of light $c$. Ignoring the factors of order unity, we can write that in terms of superdiffusion of CRs, for $M_A>1$ CRs streaming along magnetic field lines experience isotropic superdiffusion $l^2\sim (ct)^3/L$ over the scales $[l_A, L]$ for superAlfvenic turbulence and superdiffusion $l_{\bot}^2 \sim (ct)^3/L$ perpendicular to the magnetic field at the scales less than $l_A$. In the case of subAlfvenic turbulence, i.e. $M_A<1$ the superdiffusion is present perpendicular to magnetic field in the range of perpendicular scales $[l_{min}, l_{trans}]$. These results are summarized in Table~1.

Fluid viscosity may make magnetic field lines laminar at small scales. However, as discussed in Lazarian, Eyink \& Vishniac (2013), at the scales larger than the scale $l_{\|, crit} \ln({\nu/\eta})$, where $l_{\|, crit}$ is
the parallel scale of the eddies at the damping scale, $\nu$ is fluid viscosity and $\eta$ is fluid resistivity, the Richardson diffusion takes over. Fluid viscosity is important for the partially ionized gas.

 \subsection{CR diffusion along diverging magnetic field lines}
 
 The discussion above is applicable for CRs moving ballistically on scales $x$ less than the CR mean free path $\lambda_{CR}$, i.e. 
 $R<\lambda_{CR}$ and $R<L$. The dynamics of superdiffusion of CR changes when multiple scatterings happen on the scale under consideration,
 i.e. $\lambda_{CR}\ll R<L$.  

It is important to note that that the CR diffusion has two aspects. As we discussed in \S 2, an important feature of modern understanding of MHD 
turbulence is that it distinguishes the local and global mean magnetic fields. The parallel scattering should be defined in terms of the local magnetic field, as CRs trace local, rather than mean magnetic field. Scattering introduces the displacements in respect to the local magnetic field, as CR start sampling neighboring magnetic field lines and experience superdiffusion over scales $\sim \lambda_{CR}$. In other words, in terms of the mean field direction there will be displacements
arising from the diffusion of CRs along the diverging magnetic field lines and additional diffusion due to CR displacements perpendicular to the local direction of magnetic field. Below we compare the two effects.

 For the scales $\gg L$ the perpendicular diffusion coefficient is (YL08)
\be
D_{\bot, global}\approx D_{\|} M_A^4,
\label{paral_global}
\ee
where it is taken into account that for $M_A<1$ the eddies are elongated with the perpendicular dimension $\sim L M_A^2$ and crossing these eddies
involves a random walk process with the time step of  $\sim L^2/D_{\|}$. This coincides with the result for the thermal particle diffusion obtained for thermal conduction in Lazarian (2006). 

On the scales less than the $L$ a different treatment is necessary\footnote{In YL08, we attempted to introduce a diffusion coefficient that changes with the scale of the motions. We feel, however, that this way of describing superdiffusion is not useful since the superdiffusion does not obey simple Laplacian equation.} One can express the perpendicular displacement in respect to the mean magnetic field as 
\be
 l_{\bot, CR}^2 \sim \frac{(D_{\|}\delta t)^{3/2}}{27L} M_A^4,~~~M_A<1,
 \label{bot_diff1}
\ee
where we took into account that the displacement along magnetic field is governed by the parallel diffusion and therefore
\be
l_{\|, CR}^2 \approx D_{\|} \delta t.
\ee
Similarly, superAlfvenic turbulence at scales less than the Alfvenic scale $l_A$, i.e. for scales $[l_{min}, l_A]$ (see Table 1) similar arguments provide
\be
l_{\bot, CR}^{2} \approx \frac{(D_{\|} \delta t)^{3/2}}{27 L} M_A^3, ~~~M_A>1.
\label{bot_diff2}
\ee

We observe that the perpendicular displacements given by Eqs. (\ref{bot_diff1}) and (\ref{bot_diff2}) exhibit superdiffusive behavior, although the
rate of superdiffusion is reduced compared to that in the case of the Richardson diffusion. Indeed, instead of the Richardson growth in proportion
to $t^{3/2}$ for the diffusive propagation of CRs along magnetic field lines, we observe $t^{3/4}$ dependence. This is consistent with the findings in YL08. Figure \ref{MA4th_b} presents the numerical confirmation of the superdiffusive regime that corresponds to the expectations given by Eq. (\ref{bot_diff1}). 

The displacements given by Eqs. (\ref{bot_diff1}) and (\ref{bot_diff2}) are calculated in respect to the global mean magnetic field. They arise
due to the divergence of magnetic field lines as the particle diffuse tracing magnetic field lines.

In reality, particles undergo scattering and this is an additional effect that should be accounted for.
To quantify the effect of perpendicular diffusion in the presence of scattering we consider a sequence of scattering event having of CRs with the mean
free path of $\lambda_{CR}$, which is measured along the local direction of magnetic field. As the CR trace the divergent field lines they experience the effect of Richardson diffusion and the perpendicular displacement
of CR given by Eq. (\ref{bot_strong}) is
\be
l_{\bot, elem} \sim (1/3)^{3/2} L^{-1/2} \lambda^{3/2}_{CR} M_A^{2}, ~~~M_A<1,
\label{bot_elem}
\ee
where it is assumed that $l_{\bot, elem}$ is much larger than the Larmor radius $r_L$ over which the CR is being shifted perpendicular to magnetic field as a result of a scattering event.

 After each passing of the mean free path the spread is going to increase in the random walk fashion. Therefore, the total spread after $N$ scattering events
 is going to be 
 \be
 l_{\bot, cumm}^2 \sim (1/3)^{3} N \frac{\lambda_{CR}} {L} \lambda_{CR}^2 M_A^{4},~~~M_A<1 
 \label{bot_scat}
 \ee
 %To cross the largest scale strong eddies transversely, i.e. cross the distance $L M_A^2$ one requires time
 %\be
 %t_{cross, \bot}\sim 27 (L /\lambda_{CR, \|})^3 (\lambda_{CR, \|}/v_{{\|, CR}),
 %\label{crossing_bot}
 %\ee
 %where $v_{\|, CR}$ is the velocity of a CR along magnetic field. The time for
 %perpendicular crossing given by Eq. (\ref{crossing_bot}) is larger than the parallel diffusion time 
 %$t_{cross, \|}\sim (L/\lambda_{CR})^2(\lambda_{CR, \|}/v_{{\|, CR})$ and thus, unlike the case of
 %free streaming CRs do not spread  over the scale $L M_A^2$ while transversing the $L$ distance. 
 
 Due to the random walk nature of the transport in the presence of scattering one can introduce a perpendicular diffusion 
 coefficient for the diffusion perpendicular to the local magnetic field at the scales less than the perpendicular scale of the strong turbulence eddy:
 \ber
 D_{\bot, local}&\sim& \frac{R^2}{\delta t}\sim \frac{R^2}{(R^2/l_{\bot, elem}^2)\lambda_{CR}/v_{\|}}\nonumber\\
 &\sim& \frac{1}{81}\frac{\lambda_{CR}}{L} \lambda_{CR} v_{CR} M_A^4,
 \label{diff_scat_small}
 \eer
where it was used that to cross the distance $R$ due to random walk with elementary length $l_{\bot, elem}$ one requires $(R/l_{\bot, elem})^2$ steps and each step takes time $\lambda_{CR}/v_{\|}$. Finally, it was assumed that the parallel velocity of the CRs with isotropic distribution is $1/3 v_{CR}$.

The physical meaning of the diffusion described by Eq. (\ref{diff_scat_small}) is very straightforward. If CRs trace magnetic field lines in the case of free streaming that we discussed in \S 3.1, in the case of their diffusion that we discuss here, CRs, in addition, spread perpendicular to the flux tubes that they follow. This is not the usual perpendicular diffusion of CRs when every scattering event results in a CR perpendicular displacement of around the CR Larmor radius $r_L$. Due to the superdiffusion of magnetic
field lines at scales $\lambda_{CR}<L$ the CR after each scattering is being displaced by a substantially larger scale given in Eq. (\ref{bot_elem})\footnote{The Larmor radius is irrelevant for this process and the same displacements will be present for particles of different energies, including thermal particles. Thus this effect can be important for thermal conductivity of magnetized plasmas.}.

If the measurements are done in respect to the local magnetic flux tube the diffusion given by Eq. (\ref{diff_scat_small}) is the dominant effect.
However, in terms of the displacements in respect to the global magnetic field the additional diffusion presented by Eq. (\ref{diff_scat_small}) is
subdominant compared to the rate of particles separation arising from their diffusion along the divergent field lines (see Eq. \ref{paral_global}). Similar observation is also true
for the comparison of Eq. (\ref{bot_scat}) with Eq. (\ref{bot_diff1})  if one takes into account that the number of the scattering events $N\approx (D_{\|} \delta t)/\lambda_{CR}^2$.

\section{Subdiffusion of cosmic rays}

Subdiffusion is the diffusion process along magnetic field lines that undergo diffusion in space.
The subdiffusion is an ingredient for a number of models of acceleration and propagation of cosmic rays. In what follows we extend the arguments
in YL08 and define the conditions necessary for the subdiffusion to be important. 

The subdiffusion is a process widely discussed in the literature
\citep[see, e.g.][]{Kota_Jok2000, Getmantsev, Mace2000, Qin2002, Webbcompound}. If we introduced  for magnetic field lines a spatial diffusion coefficient
$D_{spat} =\delta l_{\bot}^2/\delta s$ and adopt that the transport along the magnetic field lines is diffusive, i.e. $\delta s=(D_{\|} \delta t)^{1/2}$, we 
can get the perpendicular diffusion coefficient
\be 
D_{\perp}=\left(\frac{\delta l_{\bot}}{\delta s}\right)^2D_\|=D_{spat}D_\|/\delta s=D_{spat} D_{\parallel}^{1/2} (\delta t)^{-1/2}
\label{subdiffusion}
\ee
Therefore the perpendicular transposition is 
$l_{\bot, CR}^2=D_{\perp} \delta t= D_{spat}D_{\parallel}^{1/2} (\delta t)^{1/2}$ in accordance with the findings in many papers dealing with subdiffusion.

The major implicit assumption in the reasoning above is that the particles trace back their trajectories as they are scattered backwards.
This seems possible when one considers a toy model of ``turbulence'' with random motions at a single scale that was described in \citet{RR1978} seminal paper (see more in Appendix B).  There the distance over which the 
particle trajectories get uncorrelated is comparable with the injection scale and on scales less than the corresponding
Rechester \& Rosenbluth length, the phenomenon of subdiffusion is expected. The corresponding calculations can be found
in \citet{Duffy:1995}.

The problem with this reasoning is that turbulence is not a process with one scale of random motions. We claim that this retracing requires extremely special conditions and, in fact, impossible in the presence of the Richardson diffusion of magnetic field and the retracing of CR trajectories impossible for the scales corresponding to the inertial range
of the strong turbulence. In the Appendix B we study what is expected to happen at the scales beyond the inertial range of strong MHD turbulence $[l_{min}, L_{max}]$, where $L_{max}={\rm min}(l_{trans}, L)$. There we show that the process of CR subdiffusion requires very special circumstances and therefore is very unlikely. 

\section{Modification of acceleration mechanisms in the presence of Richardson diffusion}

\subsection{Difference in parallel and perpendicular shock acceleration}

An accepted picture of cosmic ray acceleration involves a shock moving at an angle to the ordered magnetic field. In particular, two limiting cases,
the parallel and perpendicular shocks are considered, where the {\it perpendicular shock} means that the angle between the shock velocity and magnetic field is 90 degrees, while in  a {\it parallel shock} the shock velocity along magnetic field.
In the well known mechanism of shock acceleration \citep{Krymskii78, Bell78, Blandford_Ostricker}, CRs are accelerated by scattering back and forth from upstream to downstream regions. While the shock itself does not affect CRs, the compression associated with the shock induce a regular acceleration. Indeed, for the upstream CRs the downstream plasmas are approaching with the velocity $3/4 U$, where $U$ is the velocity of a strong shock. Similar effect is present for the downstream CRs crossing the shock, which results in the energy gain for the particles every time they cross the shock. This is the essence of the efficient first order Fermi acceleration in which the particle energy is increased by a factor $<\Delta p>/p=4|U_1-U_2|/(3v)$ every time it crosses the shock.

The above simple picture critically depends on the interaction of CRs with magnetic perturbations in the upstream and downstream regions. A usual textbook description of shock acceleration involves parallel shocks. There CRs streaming along the magnetic field get scattered by magnetic perturbations in the upstream and downstream regions when scattered back transverse the shock again\citep[see][]{Longairbook}. The rate of scattering limits the rate at which particles can be returned to the shock and experience another cycle of acceleration.This type of acceleration should work, but the efficiency of it is subject of debates.  

\citet{Jokipii87} noted that the rate of CRs scattering may be insufficient to explain observational data. To remedy this problem he  proposed, instead, the idea of acceleration within perpendicular shocks. 
%In Jokipii (1992) it is maintained that it is only perpendicular shocks that are able to account for the acceleration of the anomalous cosmic rays. 

The arguments in  \citet{Jokipii87} are based on the standard kinetic theory \citep[see]{Axford65} and do not take into account wandering of magnetic field
lines. \citet{Jokipii87} accepts that this is important effect but does not consider it in view of the existing uncertainties related to the process. The demonstration of the dominance of the perpendicular shocks was then very straightforward. Indeed, the time of CR acceleration from the initial 
momentum $p_{\small i}$ to $p_{\small f}$ is given by a standard expression (Forman \& Morfill 1979)
\be
\tau_{accel}=\frac{3}{V_{up}-V_{down}}\int_{p_{\small i}}^{p_{\small f}}\left( \frac{D_{up}}{V_{up}} +\frac{D_{down}}{V_{down}} \right)\frac{dp}{p},
\label{gen_sh}
\ee
where $V_{up}$, $\kappa_{up}$ and $V_{down}$, $\kappa_{down}$ are velocities and diffusion coefficients in the upstream and downstream of the shock, respectively. For the sake of simplicity while dealing with a toy problem of the shock acceleration one can accept that $D_{up}=D_{down}=D_x$, which can be presented as a combination of diffusion coefficients parallel to the mean magnetic fields $\kappa_{\|}$ and 
perpendicular to it, $D_{\bot}$
\be
D_x=D_{\|} \cos^2\theta +D_{\bot}\sin^2\theta
\label{kappa}
\ee
where $\theta$ is the angle between the magnetic field and the shock normal. 

The ratio parallel and perpendicular diffusion coefficients in kinetic theory is given by the ratio
\be
\frac{D_{\bot}}{D_{\|}}=\frac{1}{1+(\lambda_{CR}/r_L)^2}
\label{rat}
\ee
where the mean free path $\lambda_{CR}$ is larger and in many cases much larger than the CR Larmor radius $r_L$. This suggests that a perpendicular shock $\theta=\pi/2$ provides the fastest acceleration (see Eqs. (\ref{gen_sh}) and (\ref{kappa})). The possible ratio given by
Eq. (\ref{rat}) presented in \citet{Jokipii87} is $\sim (r_L/\lambda_{CR})^2$. 

The arguments above should be modified in the presence of the superdiffusion discussed in this paper. Indeed, in perpendicular direction one
expects to see the superdiffusion with the perpendicular transpositions in respect to the {\it local} magnetic field is not $r_L$ that enters in Eq. (\ref{rat}) but instead is given by Eq. (\ref{bot_elem}). Moreover, in terms of the laboratory system of reference related to the shock the transpositions in the perpendicular direction are super diffusive (see Eq. \ref{bot_diff2}). 

\subsection{Example: anomalous cosmic rays}

 Consider whether the process of superdiffusion may present a problem for accounting for a particular case when fast acceleration is required, i.e.
for the anomalous cosmic ray acceleration in the termination shock.

The anomalous cosmic rays are cosmic rays accelerated locally within the helioshphere. This problem became a hot topic in view of the failure of the Voyagers to
detect the signature of the anomalous acceleration while crossing the termination shock.
As it was argued in \cite{Jokipii92} the acceleration in the parallel shock is
too slow and only quasi-perpendicular shock has a high enough acceleration rate according to eq.(\ref{rat}). Below we discuss the conditions when the approach
in \cite{Jokipi92} is valid. 

%``the diffusive shock acceleration fails by a significant amount if one insisted on the well-established rate for quasi-parallel shocks at
%distances $\approx 70$ AU or greater''. 

To have the rates of acceleration given by Eq. (\ref{rat}) one should postulate that the magnetic field
wandering over the mean free path $\lambda_{CR}$ is less than the CR Larmor radius $r_L$. By analyzing the factors in Eq. (\ref{bot_elem}) that describes an elementary perpendicular transposition of CR at its scattering over
the mean free path $\lambda_{CR}$ on can observe that if
\be 
\frac{L}{\lambda_{CR}}> \frac{1}{27} \left(\frac{\lambda_{CR}}{r_L}\right)^2 M_A^4
\label{applic}
\ee
then the transposition of the CR due to the field wandering over the mean free path is less than the Larmor radius of the CR $r_L$. In this situation
the original treatment in   \citet{Jokipii87} are applicable. Assuming that the interstellar $M_A$ outside the termination shock is of the order of unity,
we see that the condition given by Eq. (\ref{applic}), i.e., $4.4M_A^4\left(\frac{\lambda_{CR}}{0.2 AU}\right)^3\left(\frac{B}{8\mu G}\right)^2\frac{MeV}{E_k}\frac{100Au}{L}<1$ is restrictive and may not be necessarily satisfied for the termination shock.  

Irrespectively of this particular application, our example with the termination shock illustrates the point that turbulence is known to be ubiquitous and therefore the Richardson diffusion must be considered while dealing with the problems of CR acceleration\footnote{We discuss in \S 7 how {\it small scale} turbulence in the shock precursor modifies the arguments related to the Richardson diffusion. The existence of shock precursor is associated with strong supernovae shocks and something that is not associated with the termination shock.}.

As for the acceleration of anomalous CRs in a perpendicular shock one can see from Eq. (\ref{rat}) that the efficiency of parallel and perpendicular acceleration should become comparable
in the case of the Bohm diffusion when $\lambda_{CR} \approx r_L$. Therefore we are confused by the conclusion in Jokipii (1992) that the acceleration in the parallel shock in the Bohm limit is too slow to explain the origin of the anomalous CRs, while the perpendicular shocks should do the
job of the acceleration. We find this to be in contradiction with Eq. (\ref{rat}). Thus we feel that the issue of the efficiency of the acceleration of anomalous CRs in parts of the termination shock deserves a further investigation\footnote{An alternative model of the acceleration of anomalous CRs within reconnection regions was suggested in \cite{Laz_Oph09} and was elaborated for the acceleration in collisionless reconnection in \cite{Drake10}. }.

All in all, we showed the importance of magnetic field wandering for the efficiency of perpendicular shock acceleration. Our study  testifiess that the difference between the acceleration in parallel and perpendicular shocks is reduced in the presence of the Richardson diffusion of magnetic field. To achieve much more efficient acceleration in perpendicular shocks it is required that the Alfvenic Mach number is reduced, meaning that the magnetic field should be only very weakly perturbed.  Our quantitative study in \S 6.1 supports this intuitive conclusion.

\subsection{Acceleration of particles streaming along magnetic field in shocks}

{\em The original idea}\\
An innovative idea of particle acceleration without diffusion was suggested in Jokipii \& Giacalone (2007, henceforth JG07). They considered
an adiabatic compression induced by shock interacting with free streaming particles. They discussed the idea in terms of
low-rigidity particles, e.g. electrons, but the very possibility such a streaming acceleration mechanism deserves a dedicated study. 

The idea of such an acceleration can be traced back to the analysis of \cite{Northrop63} who discussed a possibility of 
a particle acceleration in a moving magnetic loop. In the case of the shock, the physics of the process is related to the decrease
of the length of the magnetic field line due to the compression. The shock "shortens" the magnetic field lines and, as a result, 
the parallel component of the particle velocity
increases\footnote{ A similar process increases the parallel component of energetic particle velocities in reconnection
layers (de Gouveia dal Pino \& Lazarian 2005, Lazarian 2005). The turbulent reconnection layers in the LV99 model of reconnection
present 3D volumes filled with reconnecting and shrinking flux loops. This induces the First order acceleration in terms of
the angular momentum parallel to magnetic field. The process is confirmed in numerical simulations in \citet{Kowal12_acc}. A 
process for the acceleration of perpendicular component of angular momentum in the turbulent magnetic reconnection is described in
\citet{Lazarian11}, \cite{Lazarian_rev12}.}.

JG07 demonstrated that if a magnetic field
line that is nearly perpendicular to the the shock velocity is also subject to {\it random walk} excursions,
then the magnetic field line will enter the shock front many times, enabling the particles streaming along the line to get many kicks that
increase their parallel component of momentum.  

{\em Pre-existing turbulence in the pre-shock}\\
Within the JG07 mechanism the accelerated CRs are streaming and are not subject to scattering. However, turbulence is required to 
induce magnetic field wandering. Consider sub-Alfvenic or trans-Alfvenic turbulence of interstellar origin pre-existing in the upstream region. 
Assume an idealized situation when magnetic field in the upstream region is not perturbed by the shock precursor or streaming instabilities.
In many astrophysically important cases, e.g. in the case of interstellar turbulence, the injection scale is large, e.g. larger than the radius
of the supernovae shock. Thus the magnetic field lines are subject to the Richardson diffusion.

\begin{figure}
  \includegraphics[width=8cm]{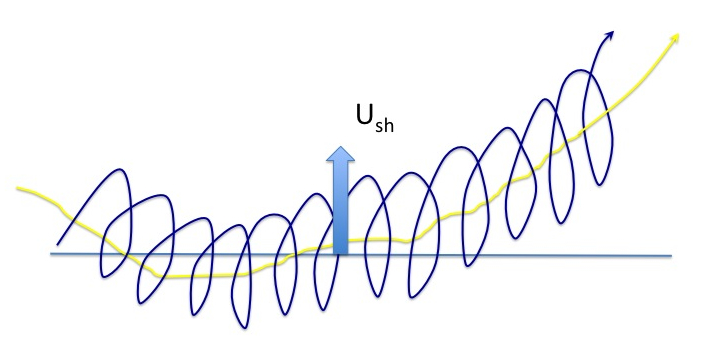}
    \caption{The particles' trajectory is diverging away from the shock front at the perpendicular shock because of Richardson diffusion on scales smaller than injection scale of turbulence $L$. If the turbulence is the pre-existing galactic turbulence with large $L$, the return of the CR while it streams along the shock is improbable. CRs can only return via scattering back as in the case of parallel shocks.}
  \lb{div_shock}
\end{figure}

The lines that cross the shock move away from the shock as  $s^{3/2}$ (see Eq. \ref{bot_strong}) and the accelerated character of this
deviation makes improbable for the magnetic field line to reenter shock  again on the scales less than the injection scale $L$. This is very
different from the random walk process considered in JG07 and therefore the acceleration of CRs this way is improbable. The fast divergence
of magnetic field from the shock region is illustrated by Figure \ref{div_shock}. 

For the same example of the supernovae shock in the interstellar medium, the driving scale of ambient pre-existing turbulence is 
$L\approx 100$~pc \citep[see][]{Chep2010}, which is larger than the size of a supernovae remnant for the Sedov phase. Indeed the turbulent magnetic field is dominant in young supernova remnants \citep{Reynolds+12}. Even if the 
injection scale of interstellar turbulence is just several pc, the return of magnetic field lines due to the random walk is not frequent enough to enable the perpendicular shocks accelerate particles with the adiabatic JG07 process. 

 We also note that streaming cosmic rays can get the increase of their perpendicular momentum through the process of drift acceleration.
 We discuss this process in \S 7 in the framework of shocks in highly chaotic small scale magnetic fields.
 In the JG07 original set-up the perpendicular acceleration is also limited by the effects of Richardson diffusion.

{\em Small scale weak turbulence as the driver for magnetic field wandering}\\
As we discussed in \S 3, the random walk of magnetic field lines is possible in the case of the weak small
scale turbulence (see Eq. \ref{diff_weak2} and Table~1). As we discussed in \S 2 the fluctuations associated with the weak Alfvenic turbulence
 are not necessarily small. If the small scale perturbations induced by CR instabilities are substantially larger than the fluctuations of strong turbulence arising from the large scale driving then the original JG07 idea works. We
consider in the Appendix C the requirements for this process to take place. We find the requirements to be rather restrictive due to the problems of generating of small scale weak turbulence in the pre-shock region.
%We the conditions for this type of acceleration can be easily met.
On the contrary,  in \S 7 we discuss how the situation is being modified in the presence of small scale superAlfvenic turbulence that is naturally generated in the shock precursor (see Beresnyak et al. 2009).

\subsection{Importance of superdiffusion}

Our analysis in the present section shows that the superdiffusive magnetic field wandering modifies the popular ideas of perpendicular shock acceleration.
While we attempt a limited quantitative study in the next section, it is clear that the Richardson diffusion of magnetic field lines presents a fundamental 
process that must be accounted for thoroughly. The quantitative description of the process presented in the present paper is intended to contribute to this task. 

 \begin{figure}
 \includegraphics[width=8cm]{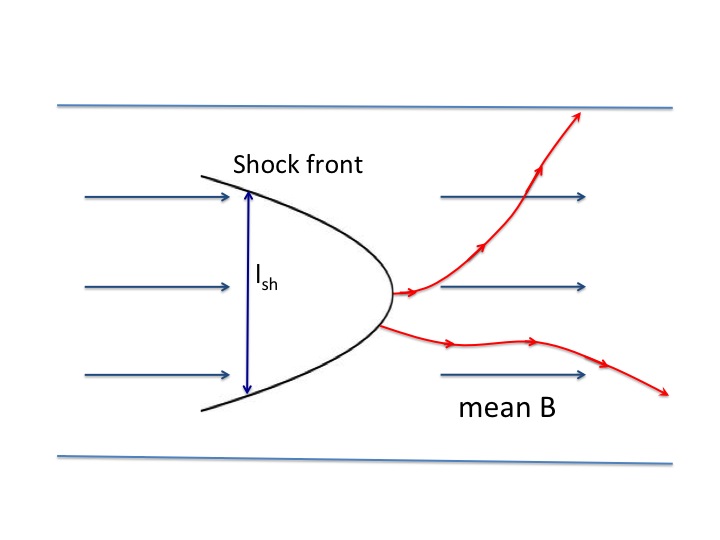}
 \caption{Streaming CRs experience Richardson diffusion and the acceleration stops once the diffusion distance becomes larger than the size of the shock $l_{sh}$.}
 % if before backscattering they get separated by the distance exceeding the diameter of
% the supernovae remnant, their return to the shock is improbable. }
 \lb{SNR}
\end{figure}

Partially ionized gas may have an appreciable viscous damping scale and on the scales less than this scale magnetic field does not exhibit superdiffusion. This is the range of scales where the earlier treatment of the differenece of perpendicular and parallel shocks in JG07 is applicable. Thus in what follows we deal with the acceleration of CRs on the scales larger than the turbulent damping scale. In the case of fully ionized gas, the Alfvenic turbulence in many cases proceeds up to the proton gyro-radius.

\section{Quantitative treatment of acceleration in the presence of superdiffusion}
We study here different regimes of shock acceleration in the presence of superdiffusion and compare them with the acceleration of usual DSA in parallel shocks.

\subsection{Acceleration time at perpendicular shock}

In diffusive shock acceleration, there is a lengthscale $l_{pen}=D/U_1$, corresponding to the distance $\sqrt{D t}$ that particles can diffuse before they are overtaken by the shock moving with speed $U_1$. Dividing the flux of particles $n v/4$ by the column density $n D/U_1$ gives the mean residence time of particles at the upstream $4D/(U_1v)$. Since the average gain of momentum per crossing is $<\Delta p>=4|U_1-U_2|p/(3v)$, the acceleration time will be \citep[see][]{Duffy:1995}
\be
t_{acc}=t_{res}\frac{p}{<\Delta p>}=\frac{3D}{|U_1-U_2|}\left(\frac{1}{U_1}+\frac{1}{U_2}\right),
\lb{tacc_dsa}
\ee
if assuming a similar expression for $t_{res}$ at the downstream.
Using a similar argument, we can get the acceleration for the case of anomalous transport. For diffusion on the scales less than injection scale, the residence time is equal to $4l_{pen}/v$, where $l_{pen}$ can be obtained by equating eq.(\ref{bot_diff1}) or eq.(\ref{bot_diff2}) with $(U_1t)^2$.
Therefore
\be
l_{pen}=\frac{D_\|^3}{U^3}\left(\frac{M_A^\zeta}{27L}\right)^2,
\lb{pen}
\ee
and
\be
t_{res}=\frac{4D_\|^3}{vU^3}\left(\frac{M_A^\zeta}{27L}\right)^2,
\lb{t_ressup}
\ee
where $\zeta=4$ for subAlfv\'enic turbulence and $\zeta=3$ for superAlfv\'enic turbulence. The acceleration time is then
\be
t_{acc}=t_{res}\frac{p}{<\Delta p>}=\frac{3D_\|^3}{|U_1-U_2|}\left(\frac{1}{U_1^3}+\frac{1}{U_2^3}\right)\left(\frac{M_A^\zeta}{27L}\right)^2,
\lb{tacc_scat}
\ee
%For the free streaming case, the acceleration time can be obtained in a similar fashion,
%e.g. using $l_{pen}$ obtained by equating Eq.(\ref{bot_strong}) or Eq.(\ref{bot_super}) with with $(U_1t)^2$, we find $t_{res}=4l_{pen}/v$, which leads to an acceleration time of
%\be
%t_{acc}=t_{res}\frac{p}{<\Delta p>}=\frac{3^7L}{M_A^\zeta v^3|U_1-U_2|}(U_1^3+U_2^3),
%\lb{tacc_fr}
%\ee
%where it was assumed again that parallel velocity of isotropic distribution $v/3$. In the case of hydrodynamical turbulence, the $M_A$ dependence disappears according to Eq.(\ref{obuch2}).

%For the case of subdiffusion \citep[see also][]{Duffy:1995}
%\be
%t_{acc}=\frac{3D_{spat}^{2/3}D_\|^{1/3}}{|U_1-U_2|}\left(\frac{1}{U_1^{1/3}}+\frac{1}{U_2^{1/3}}\right)
%\lb{tacc_sub}
%\ee

Accordingly, the ratio of acceleration time in perpendicular shock with superdiffusion (eq.\ref{tacc_scat}) and parallel  diffuse shock acceleration (eq.\ref{tacc_dsa}) is:
\ber
\frac{t_{acc,sup}}{t_{acc,DSA} }&=&\left(\frac{DM_A^{\zeta}}{27LU_1}\right)^2(r^2-r+1)\\
&\simeq&0.012M_A^{2\zeta}\left(\frac{\lambda_{CR}}{1{\rm pc}}\frac{v}{c}\frac{100{\rm pc}}{L}\frac{1000{\rm km/s}}{U_{1}}\right)^2(r^2-r+1)\nonumber
\label{accel_ratio}
\eer
$r\equiv U_1/U_2$ is the compression ratio. For strong shock with $r=4$, the above ratio is $\simeq 0.16$ much larger than $(\lambda_{CR}/r_L)^2$ as indicated by eq.(\ref{rat}). This shows that in the presence of superdiffusion, the quasi-perpendicular shocks generically have efficiencies comparable to those of quasi-parallel shocks and this presents the problem for the solution of acceleration at termination shock, for instance, as discussed in Jokipii (1992). 

The maximum energy attainable for shock acceleration can be easily obtained by equating the acceleration time with the shock expansion timescale $\sim R_{sh}/U_1$ \citep[see, e.g.][]{YLS12}. 

\subsection{Spectra of accelerated particles at perpendicular shock}
For an isotropic particle distribution, the spectrum index of shock acceleration is determined by the escape probability per cycle at the downstream $P_{esp}=4n(-\infty)u_2/[n(0)v]$, the ratio of far downstream flux to the isotropic particle flux \citep{Bell78},
\be
a=3+\frac{p}{<\Delta p>}P_{esp}.
\lb{spec_index}
\ee
Far downstream, the density distribution should relax to an unperturbed state, i.e., $n(-\infty)=Q_0/U$. 

We use the particle propagator approach in \citet{Kirk1996} to estimate the flux of particles crossing the shock where they experience superdiffusion.  Given a source function $Q_0$, the particle propagator $P(x,t)$ is defined as

\be
n(x-Ut,t)=Q_0\int_0^{\infty}dt P(x,t)
\lb{propgt}
\ee
where $n(x-Ut,t)$ is the spatial density distribution at a distance $x-Ut$ from the shock front in the upstream.

The propagator is determined by the transport property of particles. For transport with $<\Delta x^2>\propto t^\beta$, a propagator of the form below may be adopted \citep{Kirk1996}
\be
P(x,t)=t^{-\beta/2}\Phi(\frac{x}{t^{\beta/2}})
\label{kirk}
\ee
Insert it into eq.(\ref{propgt}), we can get 
$n(0)=Q_0/[U(2-\beta)]$ if we assume that the particle transport properties are the same at upstream and downstream\footnote{This assumption may not hold especially in view of different turbulence generation mechanism at preshock and postshock. We adopt the simplified model only to illustrate the impact of superdiffusion to shock acceleration.}. Put this result back into the expression for $P_{esp}$ and eq.(\ref{spec_index}), one gets
\be
a=\frac{3r}{r-1}\left(1+\frac{1-\beta}{r}\right),
\ee
Table \ref{shock} is an illustration of the effect of different transport regimes for strong shock with $r=4$. 

In the case of partially ionized media, the damping scale of turbulence is relatively large and therefore the case that the transport scale $R$ can be less than the three dimensional scale corresponding to the Rechester-Rosenbluth scale $L_{RR}$ \citep{NM01, Lazarian06} where the separations of field lines have not reached the size of the smallest eddy and they are essentially bundled together with only perpendicular displacement occurring through random walk  $<\Delta x^2>\propto \Delta z$. Below damping scale, particles are not scattered and therefore  $<\Delta x^2>\propto t$, corresponding to a normal diffusion and the momentum spectrum index of the accelerated particles is $-4$, same as the standard DSA case (see table 2).

\begin{table*}
\caption{Effect of different transport regimes to perpendicular shock acceleration}
\begin{tabular}{c|c|c|c}
\hline
\hline
& \multicolumn{2}{|c|}{$R < L$}&$R > L$\\
\hline
&{ionized medium and $R >\sqrt{L_{RR}/k_c}$ for partially ionized medium} &partially ionized medium w. $R < \sqrt{L_{RR}/k_c}$&\\
\hline     
$\beta$&3/2&1&1\\
$a$&-7/2&-4&-4\\
\hline
\hline
\multicolumn{4}{l}{\footnotesize{$\beta$ and $a$ are the power index of time for the square of displacement (see \S6.2) and 3D momentum spectrum index }}\\
\multicolumn{4}{l}{\footnotesize{of accelerated particles.}}\\
\end{tabular}\centering
\lb{shock}
\end{table*}

\subsection{Maximum energy of accelerated particles at parallel shock of finite size}

Richardson diffusion presents also a source of loss for shock with a finite spatial extent, which we illustrate on the example of a parallel shock.
Indeed, 
CRs diffusing along magnetic field lines as shown in Figure \ref{SNR} due to fast deviations of magnetic field lines may leave the part of the volume
that is going to be affected by the shock. 

Apparently for parallel shock, if particle diffuse a distance $l_{\bot, CR}$ larger than the size of a shock $l_{sh}$ within the residence time $t_{res}$, particle acceleration will cease (see Fig.\ref{SNR}). An escaping time $t_{esc}$ can be defined by equating $l_{\bot, CR}$ with $l_{sh}$. Then acceleration stops when $t_{esc}/t_{res}=1$. 

In the case of $\lambda<L$, we obtain from equating eq.(\ref{bot_diff1}) or eq.(\ref{bot_diff2}) with $4D/(U_1v)$, we get
\be
\frac{t_{esc}}{t_{res}}=\left(\frac{27Ll_{sh}^2}{M_A^\zeta}\right)^{2/3}\frac{vU_1}{4D_\|^2},
\ee
from which we get the maximum energy of accelerated particles as
\be
E_{max}=\frac{81mU_1^2}{32\lambda^4}\left(\frac{L l_{sh}^2}{M_A^\zeta}\right)^{4/3},
\ee
where $m$ is the mass of the particle. An estimate about termination shock is obtained accordingly, 
\ber
E_{max}&=&32\left(\frac{U_1}{400{\rm km/s}}\right)^2\left(\frac{L}{100{\rm Au}} M_A^\zeta \right)^\frac{4}{3}\left(\frac{l_{sh}}{90{\rm Au}}\right)^\frac{8}{3}\nonumber\\
&&\left(\frac{10Au}{\lambda}\right)^4{\rm MeV \cdot nuc}^{-1}.
\eer

\section{Fast acceleration without superdiffusion: strong turbulence generated in the shock precursor}

In the sections above we have shown how Richardson diffusion of magnetic field can modify acceleration. Its general, its effect is to decrease the 
efficiency of the acceleration in perpendicular shocks, although as we discuss in \S 6 even in the presence of superdiffusion the
perpendicular shocks may be still more efficient that in the parallel ones (see Eq. \ref{accel_ratio}). We also considered how
weak Alfvenic turbulence may help to recover the arguments on the acceleration of for freely streaming particles (\S 5.2), but noticed that 
the process is rather restrictive.  However, the Richardson diffusion of magnetic field may be neglected in the case of small scale integral scale of
magnetic perturbations. Indeed, if the mean free path of a CR $\lambda_{CR}$ that is determined by resonance scattering
is larger that the injection scale $L$. In this situation, on the scale of $\lambda_{CR}$ the turbulence induces the random walk if it is superAlfvenic perturbations making the propagation diffusive. Below we show how the arguments related to the adiabatic acceleration (see \S 5.2) can be modified for the shocks
propagating through media with stochastic field.

In the superAlfvenic turbulence the role of injection scale is played by $l_A=L/M_A^3$, as it is the scale at which magnetic fields
resist to further bending.  The diffusion coefficient in the situation $\lambda_{CR} \gg l_A$, where $\lambda_{CR}$ is calculated through
scattering calculations is (Lazarian 2006, YL08)
\be
D\approx 1/3 l_{cs} v_{CR}.
\label{kap}
\ee
where $l_{cs}$ is the integral correlation scale of magnetic perturbations, that in the case of superAlfvenic turbulence is equal to $l_A$. 

The acceleration in the case of $l_{cs}\ll \lambda_{CR}$ has its own features. First of all, one can clearly see that the process of
increasing the parallel to local magnetic field component of CR momentum considered in JG07 is applicable to this situation. If the 
 time for the magnetic eddy $l_{cs}$ is $t_{conv}\approx l_{cs}/U$, where $U$ is the shock velocity, then the number of eddies sampled by
 a CR during the shock compression time is 
 \be
 t_{conv} \frac{v_{CR}}{l_{cs}}=\frac{v_{CR}}{U}
 \lb{req}
 \ee
which is $\gg 1$ for non-relativistic shocks that we consider. This proves that the acceleration is expected to be efficient. The corresponding spectrum of the accelerated particles is \citep[see][]{Jok_Giac07}:

\begin{eqnarray}
f(x,p_\|)=\cases{\propto p_\|^{-a}{\rm exp}(U_1x/D) &  for upstream\cr
\propto p_\|^{-a} &  for downstream\cr},
\label{hbcoll}
\end{eqnarray}
where $a=r/(rM_{A,up}^2-M_{A,down}^2)$.  For the diffusion coefficient D, we should use Eq.(\ref{kap}) instead of the original one in JG07 for the fast acceleration we discuss here.

However, we claim that not only parallel component of the CR will increase due to the compression induced by a shock. The processes of
scatter free drift acceleration in perpendicular shocks \citep{Armstrong_Decker1979, Pesses1979} should be important for the parts of the shock where the the magnetic field is parallel to the shock front. 
Indeed, consider a process of the interaction of a CR moving along the loop of the size $l_{cs}$ and interacting with a magnetic 
mirror created by an adjacent loop moving due to the compression. Naturally, this process increases the perpendicular
momentum of the CR. As a result we expect both parallel and perpendicular components of the CR momentum to increase. 

The maximal
energy available through this process is determined by the condition that the Larmor radius of the CR $r_L$ is equal to the integral correlation
scale of the magnetic field $l_{cs}$, i.e.
\be
E_{max}\approx \frac{l_{cs}B(\mu G)}{5\times 10^{11}{\rm cm}}{\rm GeV},
\lb{max}
\ee
For CRs of energies larger than given by Eq. (\ref{max}) the acceleration proceeds in the diffusive regime as the turbulence on the scale less than $r_L$ and $\delta B >B$ induces stochastic behavior of energetic particles. 

The acceleration rate for the drift acceleration is \citep{Kota79}
\be
\frac{\dot{p}_\bot}{p}=-\frac{1}{3}\nabla\cdot U_\bot
\ee
and the acceleration rate for the parallel momentum increase due to the decrease of magnetic loop is

\be
\frac{\dot{p_\|}}{p}=-\frac{\partial(U<(B_x/B)^2>)}{\partial x}
\ee
In the case of strong turbulence, we see that the rates of the two are comparable.

The generation of small scale entangled field in front of the shock can be done either through
either the non-resonant instability proposed by \citet{Bell2004} or
 by turbulent dynamo in the  precursor as is suggested in \citet[][henceforth BJL09]{BJL09}. Both mechanisms were proposed to explain the 
 acceleration of high energy CRs, but here we are interested in the chaotic small-scale structure of magnetic field that is produced by these processes.
 The injection scale for the chaotic magnetic fields is much less than the spatial extend of a supernovae shock and therefore particles, e.g. low
 rigidity particles, streaming along magnetic field lines may enter and cross the shock many times. 
 
 Consider the BJL09 process. The precursor is an accepted part of the picture of CR acceleration. It is created by the accelerated CRs streaming ahead of the shock and reflected back \citep{Diamond_Makov}. BJL09 predicts that when the precursor interacts with the inhomogeneities of the turbulent density pre-existing in the upstream, this generates vorticity and turbulent motions that in their turn generate magnetic fields in the precursor via
 turbulent dynamo \citep[see][]{CVBL08}. The characteristic
 scale of the largest eddies is limited by the thickness of the precursor. The latter may be much larger than the magnetic field structures at $l_A$ scale.
 The details depend on the properties of the precursor and the density inhomogeneities pre-existing turbulence (see BJL09 for details).
 
 Figure \ref{precursor} shows the acceleration that the shock induces for the particles streaming along magnetic field entangled on the scale
 $L$ much smaller than $\lambda_{CR}$. In view of parallel acceleration of the acceleration in terms of the particle momentum parallel to magnetic
 field, the process is similar to that discussed in JG07, but in the presence of chaotic field entangled at small scales, there is no difference between
 the parallel and perpendicular shocks. In fact, the process can be referred to as ``shock acceleration in entangled magnetic field''.

\begin{figure}
 \includegraphics[width=8cm]{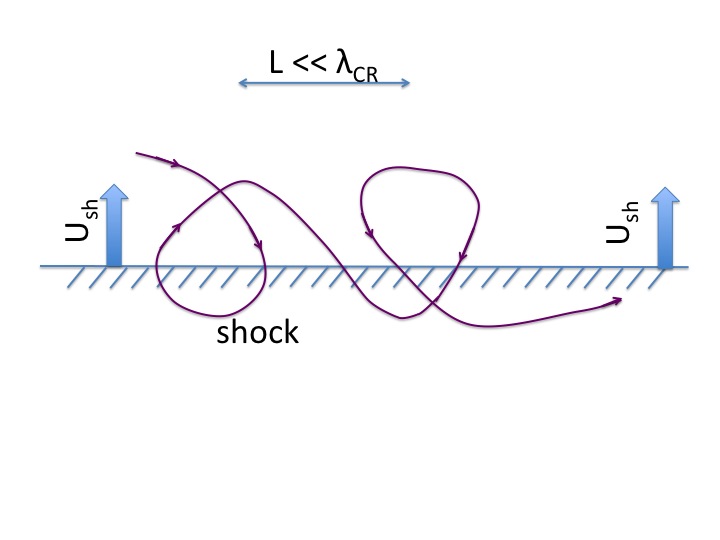}
 \caption{Schematics of the CRs acceleration in the process of free streaming for $\lambda_{CR}$ much larger than the scale $L$ at which the magnetic
 field is entangled. In general, the magnetic field structure in the pre-shock and post-shock may differ, but we disregard this complication.}
  \lb{precursor}
\end{figure}

\section{Discussion}

\subsection{Superdiffusion and CR acceleration in entangled field}

Our study shows that while the diffusive shock acceleration is applicable to parallel shocks, the effects arising from the fact that 
Richardson explosive diffusion of magnetic field lines (LV99, ELV11) induce super diffusive behavior of CRs. Fast deviations of magnetic 
field lines from the mean direction of magnetic field allow CRs to diffuse in the direction of the shock velocity faster than in the case of a 
classical perpendicular shock. This naturally makes the parallel and perpendicular shock acceleration comparable.

The issues related to superdiffusion are not present for the scale larger than the scale of entanglement. In the paper
we discuss that the scale of magnetic field entanglement can substitute the mean free path of the CR, decreasing CRs diffusivity and making
the acceleration more efficient. Moreover, the CRs streaming along entangled magnetic field lines can experience adiabatic acceleration in
the shock. All this makes the acceleration in "shocks in entangled field" a subject that deserves further careful investigation.

We note, that the entangled magnetic fields are being naturally produced in the pre shocked regions through the interaction of the precursor with
the density inhomogeneities existing in the ambient media. They also can be generated by various instabilities induced by CRs. Future research should
shed more light on the nature of the entangled magnetic field in the pre shock and post shock regions in different environments and enable researchers to make detailed quantitative calculations. 

\subsection{Perpendicular diffusion of CR and reconnection of magnetic field lines}

A major subject that this paper deals with are the implications of the Richardon diffusion for the CR propagation and acceleration.
The issues of magnetic field wandering corresponding to the Richardson diffusion have been the focus of the recent discussion of
problems of magnetic reconnection and flux freezing violation in turbulent media (LV99, ELV11, see Lazarian, Eyink \& Vishniac 2013 for a review). Both 
the theoretical predictions of fast magnetic reconnection and the violation of flux freezing have been supported by numerical simulations \citep{Kowal09_rec, Kowal12_rec, EV13_nature}. The natural question is to what extend the textbook treatment of magnetic fields in fluids in conducting fluids is valid in the presence of
the violation of flux freezing.

The answer to the question above depend on the problem that one is dealing with. We claim that for CRs the complicated dynamics of ever-changing reconnecting magnetic field lines is not essential\footnote{The static magnetic field assumption is not true for relativistic environments, but we do not deal with such cases in this paper. For other situations when superdiffusion is important, the dynamics of magnetic field lines can be essential. For instance, for the process of the removal of magnetic flux from molecular clouds and disks via reconnection diffusion \citep[see][]{LE12}, the dynamics of magnetic field is essential. In other processes, e.g.
thermal conduction of plasma in galaxy clusters, both the motion of electrons along the fast diverging magnetic field lines and the dynamics of magnetized eddies are important (see Lazarian 2006). }. A CR samples the instantaneous complicated and stochastic field line
that it is moving along. The existence of such a field line is due to the ability of the field lines to reconnect rather than forming small scale knots at the intersection of field lines. In fact, this provides the necessary condition for the complicated field line wandering required in all the models of efficient perpendicular diffusion.

%Our paper explains the superdiffusive nature of the particles moving along magnetic field lines at scales less than the injection scale of the turbulence. As the injection scale of the turbulence in the ISM is of the order of $\sim 100$ pc, the paper addresses the issues of CR dynamics over appreciable scales if spiral galaxies are concerned. 

\subsection{Different regimes of superdiffusion}

If cosmic rays are moving ballistically, their perpendicular displacement grows as $s^{3/2}$, while on the scales larger than the mean free path of the cosmic ray,
i.e. $\lambda_{CR}<s<L$ the perpendicular displacements grows as $s^{3/4}$. The mean free path is determined by the turbulent scattering of CRs. Yan \& Lazarian (2002, 2004) identified the fast MHD modes as the major CR scattering agent. Therefore the efficiency of scattering depends not only on the level of
turbulence, but on the fraction of turbulent energy associated with fast modes. The coupling and energy transfer between Alfven, slow and fast modes is suppressed (Cho \& Lazarian 2002, 2003).
% and in the presence of collisionless damping of fast modes the mean free path of a CR may become large\footnote{The issue of the efficiency of collisionless damping does require further studies in turbulent compressible plasmas. The effective mean free path of collisionless particles may be substantially reduced in the presence of instabilities induced by turbulence (see Lazarian \& Beresnyak 2007, Yan \& Lazarian 2011, Santos-Lima et al. 2013)}. 
%This was used
%in Brunetti \& Lazarian (2011) to claim that the collisionless damping may be substantially reduced in turbulent plasmas (see also . }.

The two regimes of superdiffusion are confirmed by numerical simulations in Xu \& Yan (2013) where data cubes obtained by MHD turbulence simulations were
used to study particle propagation. We note, that superdiffusion cannot be described by a diffusion coefficient. We also note that the while talking about perpendicular diffusion of CRs one should specify whether the diffusion is considered in respect to the mean field or to the local field. We showed in this paper (\S3.3) that the results can be very different in these two cases. 

%\subsection{Subdiffusion of CRs}

%Our study supports the notion in YL08 that the process of subdiffusion requires very special circumstances. The process requires that the CRs
%retrace their trajectories along magnetic field lines. In the presence of the Richardson diffusion of magnetic field lines, the lines separated over
%the distance less than the Larmor radius of a CR, get separated by a substantial distance in an explosive fashion. This makes the retracing impossible
%for CRs propagating in strong MHD turbulence of superAlfvenic turbulence (see Table~1). In other regimes, as we discussed in the paper, subdiffusion
%also requires special conditions to be satisfied. Thus, unlike superdiffusion, subdiffusion is a very unlikely process. 

\subsection{Acceleration and superdiffusion of charged dust}

Charged dust particles behave similar to CRs, but their velocities and the charge to mass ratios are enormously different. They, unlike cosmic rays, can also be
accelerated by hydromagnetic drag \citep{LY02}. The theories of dust acceleration by magnetic fluctuations and hydrodynamic turbulent motions 
have been developed in a number of papers \citep{YL03, YLD04, Yan09, HLS12}. 
%The acceleration of charged dust by shocks was considered in \cite{}Giacalone \& Jokipii (2010).  
The stochastic acceleration of dust particles by turbulence can pre-accelerate them prior to the acceleration in shocks. Similar to the case of CRs, for dust acceleration and propagation the effects of superdiffusion can be important though the scale range that superdiffusion applies is smaller than the case for CRs because of the larger radii of dust grains.

\subsection{CR acceleration in shocks and sites of turbulent reconnection}

The process of turbulent reconnection induces the First order Fermi CR acceleration \citep{DL05, Lazarian05, Kowal11_acc, Kowal12_acc}.
CRs in the reconnection region move along the contracting loops gaining energy in a regular way. Unlike the case of shock acceleration, the distribution of particles
is required to be anisotropic for the process to proceed efficiently\footnote{The effects of compression due to reconnection are subdominant
and the calculations in \cite{CL06} show that  the acceleration efficiency in incompressible flows decreases dramatically in the presence
of efficient scattering.}.

The superdiffusion of CRs introduces an additional channel of losses from the reconnection region, similar to the process that we described in \S 6.3
These losses are expected to increase with the increase of the Alfven Mach number of turbulence within the reconnecting magnetic fluxes. A more
detailed discussion of the effect requires further studies and beyond the scope of the present paper. 

In the paper we noticed a possible analogy between CRs acceleration in reconnection sites and shocks. Indeed, as we discussed in \S 7 CRs freely streaming along entangled magnetic field  at small scales can experience fast 
acceleration in the shocks. The parallel component of CR momentum increases due to the effective decrease of a magnetic loop subjected to
the shock compression. In reconnection sites, magnetic reconnection also decreases the length of magnetic field lines inducing parallel acceleration.
A drift acceleration is present for the perpendicular component in both cases as well. Both processes of acceleration require further studies. 

%\subsection{Dynamics of super diffusive magnetic field lines}

%\subsection{Observational tracing of magnetic fields}

%In view of the big implications of superdiffusion for CR physics, it is advantageous to study the structure of magnetic fields. Such tracing is difficult,
%however, with synchrotron, where localized measurements of synchrotron polarization are rarely possible. Aligned dust
%provides a slightly better way to study local magnetic field structure, as variations of grain alignment  (see \citealt{Lazarian07rev} for a review) can be employed.
%One of the most promising ways to study 3D structure of magnetic field is to use atomic alignment \citep{YLfine, YLhyf, YLHanle, YL12_rev, SY13} or, in the case of solar wind, in situ measurements.   

\section{Summary}

In the paper above we considered MHD turbulence driven isotropically at the injection scale $L$ and considered the propagation of
cosmic rays at different regimes. We showed that 
\begin{itemize}
\item At the scales less than the turbulence injection scale the perpendicular
dynamics of cosmic rays is super diffusive, the separation between CRs grows faster than square root of time.
This is not related to the hypothetical Levi flight behavior, but is due to the well-established divergence of magnetic field lines related to the process
of the Richardson diffusion. As the injection scales of turbulence in galaxies may be equal to larger than a hundred parsec, the superdiffusion
must be accounted in the models of cosmic ray propagation and acceleration. 

\item The superdiffusion in the case of ballistic propagation, i.e. on
the scales less than the mean free path of a CR, induces the CR separation that is similar to the separation of magnetic field lines and 
grows as time to the power of $3/2$. On the scales larger
than the mean free path, the CR separation grows as time to the power of $3/4$. At scales larger than the turbulent injection scale cosmic rays
exhibit diffusive behavior.

\item In contrast to superdiffusion, the subdiffusion is extremely special improbable phenomenon that, as we discuss in the paper, require very special conditions.

\item The superdiffusion changes the properties of the acceleration of CR acceleration in shocks. In particular, the superdiffusion {\em diminishes}
the difference that is present between the parallel and perpendicular shocks. 

\item The process of CR acceleration for the shocks with the precursor with small scale magnetic field may be efficient for the CRs ballistically
moving along turbulent magnetic field lines. 

\end{itemize}

\begin{acknowledgements}
We are grateful to the referee for his/her comments. We thank Greg Eyink for valuable comments and Randy Jokipii for useful exchanges. AL acknowledges the support of the NSF grant AST-1212096, the NASA grant NNX09AH78G, the Vilas Associate Award as well as the support of the NSF Center for Magnetic Self-Organization. In addition, AL thanks the International Institute of Physics (Natal, Brazil) for its hospitality. HY acknowledges the support by NSFC grant AST-11073004 and visiting fellowship at U Montpellier and Observatoire Midi-Pyren\'ees in Toulouse. 
\end{acknowledgements}

\appendix

\section{A: Quantitative description of the spatial divergence of magnetic field lines}

To describe the divergence of magnetic field lines it is convenient to define a ``2-line diffusivity'' (ELV11)
\be 
D_{ij}^B(\Bell) = \int_{-\infty}^0 ds\,\langle\delta \hat{b}_i(\Bell,0)
\delta\hat{b}_j(\Bell,s)\rangle, 
\lb{DB} 
\ee
where $\delta\hat{b}_i(\Bell,s)=\hat{b}_i(\boxi'(s))-\hat{b}_i(\boxi(s))$
with $\Bell=\boxi'(0)-\boxi(0)$. Thus the corresponding diffusion tensor is
\be 
{{d}\over{ds}}\langle \ell_i(s)\ell_j(s)\rangle=\langle D_{ij}^B(\ell)\rangle. 
\ee

In agreement with LV99 theory of magnetic field wandering the integrand in (\ref{DB}) for 
field-perpendicular increments  can be presented as  
\be  
\langle\delta \hat{b}_\perp(\Bell,0)\delta\hat{b}_\perp(\Bell,s)\rangle \sim
    \frac{\delta u^2(\ell)}{v_A^2}{{\rm Re}}\left[e^{is/\ell_\|-|s|/\lambda(\ell)}\right], 
 \lb{dBdBs} 
 \ee
where $|\delta \hat{b}_\perp(\ell)|\sim \delta B(\ell)/B_0\sim \delta u(\ell)/v_A.$ 

Consider the two terms in the $s$-dependent part. It is easy to see that the
 imaginary exponent represents  Alfvenic
 oscillations with the scale $\ell_\|$, where $\ell_\|,$ is the parallel length-scale of GS95 eddies
with the parallel and perpendicular dimensions  $\ell_\|$ and $\ell_\perp$, respectively (see \S 2).
 
  The other term includes a correlation length $\lambda(\ell)$
of tangent-vector increments along the field line. This term describes an exponential decay of
 correlations along the field. In LV99 theory $\lambda(\ell)$  corresponds to the distance traveled by an
Alfvenic perturbation along the field-line with velocity $v_A$ during the cascading
  time for turubulence at scale $\ell_\perp.$ In other words, 
\be  
  \lambda(\ell) = v_A\tau_\ell=v_A \frac{\delta u^2_\ell}{\varepsilon}. 
  \lb{Lambda}
\ee

As (\ref{DB}) can be written as  
\be
 D_{ij}^B(\Bell)\sim \delta \hat{b}_i(\Bell)\delta \hat{b}_j(\Bell)s_{int}(\Bell),
 \lb{DB2} 
 \ee
where $s_{int}(\Bell)$ is an integral correlation length of the increment in the tangent 
vector along the lines. One should deal with the properties of $s_{int}(\Bell)$ in order
 to describe the line separation.

Integrating (\ref{dBdBs}) in $s$ gives
the following result
\be 
s_{int}(\ell) \sim \frac{1/\lambda(\ell)}{1/\lambda^2(\ell) + 1/\ell_\|^2}\sim 
        \frac{\ell_\|^2}{\lambda(\ell)}= \frac{\varepsilon}{v_A} 
        \frac{\ell_\|^2}{\delta u^2_\ell} \lb{sint},
\ee
for $\lambda(\ell)\geq \ell_\|.$ Substituting into (\ref{DB2}) one obtains
\be 
D^B_\perp(\ell) \sim \frac{\varepsilon\ell_\|^2}{v_A^3}=\frac{\ell_\|^2}{L}M_A^4, 
\lb{DB3} 
\ee
where the factor of $\delta u^2(\ell)$ got cancelled. 

Consider different regimes of turbulence.
In the strong GS95 turbulence regime, the condition of critical balance requires that  $\lambda(\ell)
\sim \ell_\|$ and thus, from Eq. (\ref{sint}), $s_{int}(\ell)\sim\ell_\|.$ Thus one can write 
\be 
{{d}\over{ds}}\ell_\perp^2\sim D_\perp^B(\ell)\sim (\delta u_\ell/v_A)^2 \ell_\| \sim L\left({{\ell_\perp}\over{L}}\right)^{4/3}
M_A^{4/3}, 
\label{a_diff_strong}
\ee
where we have substituted from Eqs. (\ref{Lambda}),(\ref{vl}) for $\ell_\|$ and $\delta u_\ell/v_A.$.

In the weak turbulence regime one should use 
$\ell_\|=L$ being a constant, which when substituted into (\ref{DB3}) gives 
\be  
{{d}\over{ds}}\ell_\perp^2\sim D_\perp^B(\ell) \sim L M_A^4,
\label{a_diff_weak}
\ee
and this is solved to give the result  corresponding to the ordinary diffusion 
\be
\ell_\perp^2\sim sLM_A^4.
\ee
These results are used in the main part of the paper.

\section{B: Subdiffusion as an unlikely process}

Consider first the scales less that the viscous dissipation scale. For magnetic field to be present at these scales the viscosity of the fluid should be larger
than the resistivity, i.e. the fluid should have high Prandtl number. At scales less than $l_{min}$ magnetic field is stirred by the turbulence of larger eddies. The most important fastest steering arises from the marginally damped eddies of the size $l_{min}$. These are still Alfvenic anisotropic eddies
and therefore it is appropriate to characterize them by two distinct scales, the perpendicular scale $l_{\bot, min}$ and $l_{\|, min}$,$l_{\bot, min}<l_{\|, min}$, The eddy type motions correspond to motions perpendicular to the local direction of magnetic field, thus $l_{\bot, min}$ can be identified with
the single scale of driving in the  Rechester \& Rosenbluth theory \citep{NM01, Lazarian06}. In that theory it was calculated the path length of the particles for them to get separated by the driving scale and therefore lose the ability to retrace their trajectories. The corresponding Rechester \& Rosembluth scale is given by the expression in YL08:
\be
L_{RR}=l_{\|, min}\ln(l_{\bot, min}/r_{CR}),
\label{RR}
\ee
where the only difference from the corresponding expression in \cite{Lazarian06} is its using the Larmor radius of CR, $r_{CR}$, instead of the thermal electron Larmor radius in the latter work. Due to the slow growth of the logarithm, $L_{RR}$ in Eq. (\ref{RR}) is of the order of $l_{\|, min}$. 
Therefore it follows from Eq.(\ref{RR}) that there can be diffusion of magnetic field lines with a spatial diffusion coefficient $D_{spat}=\delta l_\bot^2/\delta s$. However, parallel transport of particles below $L_{RR}$ is unlikely to be diffusive lacking of perturbations to scatter. This makes the subdiffusion process rather exotic and improbable below $l_{min}$

Consider now whether the process of subdiffusion is possible for scales larger than $L_{max}$. Naturally, if one considers CRs with the Larmor
radius $r_L$ less than $L_{max}$, such CRs will trace the super diffusing magnetic field lines and the retracing with such CRs is impossible. However, if
$r_L$ is larger than $L_{max}$ the cosmic ray would trace large scale magnetic field undergoing the random walk (see Figure \ref{lazarianetal} for large scales). For instance, for $M_A<1$ a cosmic ray with $r_L>l_{trans}$ would interact with perturbations arising from weak turbulence while moving along the mean magnetic field. Such perturbations according to Eq. (\ref{bot_weak}) produce random walk spacial diffusion with the step $L$, while the CR may can also get scattered by magnetic perturbations along the mean field. In this case we do have a case of subdiffusion. The scattering of the CR may arise from, e.g. cosmic ray instabilities, e.g. from a streaming instability. 

All in all, due to the Richardson diffusion of magnetic field lines, the process of subdiffusion is not possible over the range of scales corresponding to the inertial range of turbulence. The subdiffusion is extremely unlikely for scales less than the scale of viscosity damped eddies, but still possible for the CRs with Larmor radii larger than the injection scale of the strong turbulence, e.g. the scale $l_{trans}$ for subAlfvenic turbulence. In the latter case some additional mechanism should provide the source of perturbations to induce cosmic ray scattering. 

\section{C: Weak turbulence and acceleration of free streaming particles in a shock}

The case that the original idea in \citet{Jokipii87} works if the small scale {\it weak} Alfvenic turbulence is generated with the injection scale much less than the scale of the system. Such weak turbulence would induce random walk displacements according the Eq. (\ref{bot_weak}) and can provide multiple crossings of the shock front. Cosmic ray instabilities, e.g. cosmic ray streaming, can produce waves, that while being scattered back through the parametric instability or through the reflection from density inhomogeneities pre-existing in the turbulent pre-shock environment produce weak turbulence at small scales\footnote{We note that to have Alfvenic turbulence rather than Alfvenic waves, the colliding Alfvenic packets should move in the opposite directions.}. As we discussed in \S 2 the weak and strong MHD turbulence do not reflect the amplitude of Alfvenic perturbations, but only the strength of non-linear interactions. Therefore the amplitude of magnetic perturbations arising from weak turbulence may substantially exceed, on the small scales that we are dealing with, the amplitude of perturbations arising from the large scale strong Alfvenic furbulence. 

An interesting feature of this scenario is that the weak turbulence produced by CRs is competing with the strong pre-existent turbulence in the
upstream. The latter induces Richardson explosive separation of magnetic field lines according to Eq.(\ref{bot_strong}) and this way decreases the possibility of a magnetic field line to re-enter the shock, while the former induces random walk wandering of magnetic field lines according to
Eq.(\ref{bot_weak}) thus helping magnetic field line to re-enter the shock many times. Note that the injection scales in 
Eqs.(\ref{bot_strong}) and (\ref{bot_weak}) are very different. For the strong pre-existing turbulence the injection scale is determined, e.g. by
the large scale steering of the interstellar gas, while for weak turbulence generated by CRs the injection scale is the scale of the perturbations created by
the instabilities may be of the order of the order of gyro radius $r_L$. 

The above scenario has obvious limitations. CRs streaming instability may be the primary source of perturbations. However, if a magnetic field line enters many times the shock this is creates many  points of injection over the length of the magnetic field line which limits imbalance in the flow of oppositely moving cosmic rays. Potentially, gyro resonance instability (see Kulsrud 2007) arising from the anisotropic distribution of accelerated particles of the momentum space can also produce waves and thus generate weak turbulence. We do not attempt to quantify these possibilities here, but want
to mention that in this scenario a possibility that CRs of low energy and high energies are getting different types of acceleration. The low energy ones create the instability and get scattered and reflected back as in a usual picture of parallel shock acceleration and at the same time create weak turbulence that induces magnetic field random walk. The instability for higher energy particles may experience turbulence suppresses the streaming instability by the ambient turbulence\footnote{This is another process of a competition
between the pre-existing turbulence in the pre-shock and the turbulence generated by cosmic rays.} \citep{YL02, FG04, BL_wave}. Therefore such particles can stream freely along the the magnetic field lines, which, due to weak turbulence, undergo random walk and exhibit multiple entries of the shock as it is suggested in \citet{Jokipii87}.

\bibliographystyle{apj.bst}
\bibliography{yan}

\clearpage

\end{document}